\documentclass{elsart}
\usepackage{epsfig}
\usepackage{amsmath}
\usepackage{color}
\usepackage{amssymb}
\usepackage{graphicx}
\usepackage{subfigure}

\newcommand \be{\begin{equation}}
\newcommand \ba{\begin{eqnarray}}
\newcommand \ee{\end{equation}}
\newcommand \ea{\end{eqnarray}}

\begin{document}
\begin{frontmatter}
\title{Bubble, Critical Zone\\and the Crash of Royal Ahold}
\author[NL]{\small{Gerrit Broekstra}},
\author[iggp,ess,nice]{\small{Didier Sornette}\thanksref{EM}},
\author[iggp]{\small{Wei-Xing Zhou}}
\address[NL]{Organization Behavior and Systems Sciences,
Nyenrode University, Breukelen, The Netherlands}
\address[iggp]{Institute of Geophysics and Planetary Physics,
University of California, Los Angeles, CA 90095}
\address[ess]{Department of Earth and Space Sciences,
University of California, Los Angeles, CA 90095}
\address[nice]{Laboratoire de Physique de la Mati\`ere Condens\'ee,
CNRS UMR 6622 and Universit\'e de Nice-Sophia Antipolis, 06108 Nice Cedex 2, France}
\thanks[EM]{Corresponding author. Department of Earth and Space
Sciences and Institute of Geophysics and Planetary Physics,
University of California, Los Angeles, CA 90095-1567, USA. Tel:
+1-310-825-2863; Fax: +1-310-206-3051. {\it E-mail address:}\/
sornette@moho.ess.ucla.edu (D. Sornette)\\
http://www.ess.ucla.edu/faculty/sornette/}

\begin{abstract}

Our analysis of financial data, in terms of super-exponential
growth, suggests that the seed of the 2002/03 crisis of the Dutch
supermarket giant AHOLD was planted in 1996. It became quite
visible in 1999 when the post-bubble destabilization regime was
well-developed and acted as the precursor of an inevitable
collapse fueled by raising expectations of investors to maintain
strong herding pressures. We have adapted Weidlich's theory of
opinion formation to describe the formation of buy or sell
decisions among investors, based on a competition between the
mechanisms of herding and of personal opinion opposing the herd.
Among four typical patterns of stock price evolution, we have
identified a ``critical zone'' in the model characterized by a
strong sensitivity of the price trajectory on the herding and
personal inclination parameters. The critical zone describes the
maturation of a systemic instability forewarning of an inevitable
crash. Classification and recognition of the spontaneous emergence
of patterns of stock market evolution based on Weidlich's theory
of complex systems, and in particular our discovery of the
post-bubble destabilization regime which acts as a precursor to a
subsequent crash or antibubble, not only presents the possibility
of developing early warning signals but also suggests to top
management ways of dealing with the coming crisis.

\end{abstract}


\end{frontmatter}

\typeout{SET RUN AUTHOR to \@runauthor}

\section{Introduction} \label{s1:introduction}

The 21st century opened with a deep confidence crisis in the financial
markets caused by unprecedented corporate scandals both in the USA
(Enron, Worldcom) and in Europe (Ahold, Parmalat).  Financial
authorities and investors would greatly benefit from a systematic
analysis of publicly available corporate financial performance data such
as sales and earnings and, particularly, stock price dynamics of
high-growth companies that would enable them to detect early warning signals
of impending problems in bullish times. It is our purpose to present
a study showing the relationship between these variables and a model
to understand the origin of crises in individual companies.

In the wake of the worldwide stock markets bubbles followed by
crashes or `antibubbles' of the second half of the 1990s and early
2000s, the analysis of such critical phenomena in aggregate stock
markets has been intensified; see for instance (Abreu, 2001;
Richardson and Ofek, 2001; Visano, 2002; Caballero and Hammour,
2002; Bohl, 2003; Brooks and Katsaris, 2003; Siegel, 2003;
Scheinkman and Xiong, 2003; Griffin et al., 2003; Brunnermeier and
Nagel, 2003; Chari and Kehoe, 2003; Kaizoji, 2004; Engsted and
Tanggaard, 2004). A series of synthesis papers (including
Kindleberger (2000), Shefrin (2000), Shiller (2000), Shleifer
(2000), Sornette (2003)) have pointed out the role of collective
behaviors, such as herding and optimism feedback on itself, in the
development of bubbles in aggregate markets. Far less attention
has been paid to the fate of individual companies during these
bubbles (see however (Johansen and Sornette, 2000) and (Lamdin,
2002)) and particularly those high-growth companies that couldn't
resist to manipulate their earnings to influence investors'
behavior eventually ending up with cutting their own flesh. That a
propensity of earnings manipulation during bullish times is not
unusual was recently emphasized from research in the Chinese stock
market (Jiao, 2003).

The present work has two main goals. First, we study the recent
bubble and its aftermath of the Ahold company. Our purpose is to
extend previous works on bubbles and crashes that were essentially
performed on aggregates, such as indices or major currencies. We
particularly refer to (Johansen et al., 1999; 2000; Sornette and
Johansen, 2001; Sornette and Zhou, 2002; Zhou and Sornette, 2003;
Sornette, 2003; Johansen and Sornette, 2004), where a quantitative
framework has been developed to test for the presence of
speculative bubbles and to predict their termination, often in the
form of crashes. In a nutshell, these works propose that
speculative bubbles reflect the interplay between preponderant
positive feedbacks modulated by intermittent negative feedbacks,
leading to characteristic ``log-periodic power law'' (LPPL)
signatures. These works have tested the LPPL signatures on major
financial aggregates. Johansen and Sornette (2000)
have previously performed a rapid analysis of the price of the
shares of IBM and of Procter \& Gamble and showed the presence of
a speculative bubble preceding the crash in both cases.
Professionals have used this methodology to develop trading
techniques on individual companies (private communications) but we
are not aware of other published works on other individual
companies. To attain sufficient depth, we focus in this article on
one particular case, Royal Ahold, which is of interest to both
sides of the Atlantic, because a relatively large share of the
sales of one of the world's largest supermarket chains occurs in
both Europe, and the Netherlands in particular, and in the USA. We
address the questions to what degree bubbles can occur on
individual companies, how they are linked to objective variables
characterizing the company (such as sales and earnings), how they
are coupled with the dynamics of related indices.

Our second goal is to introduce the new concept of a ``critical
zone'' characterizing the termination of a bubble and the
transition to another regime, be it via a crash, a correction or
simply a plateau. Johansen et al. (1999; 2000) have developed a
rational expectation (RE) model of bubbles and crashes
incorporating herding behavior, which shows that the bubble
termination time is the instant at which a crash is the most
probable (but is still not certain). In this class of RE models, a
crash can occur at any time but is more and more probable as the
bubble develops. In addition, there is a finite probability that
no crash occurs, that is, that the bubble ends smoothly. However,
the question remains open as to how the bubble transitions to a
different phase after its demise. Sornette and Zhou (2004) have
shown that the LPPL theory can be used in a general pattern
recognition method with efficient generalizing properties to
detect times of changes of regime which are not necessarily or
immediately associated with a crash (that is a sharp price drop of
say more than 15\% occurring over a short time interval of say
less than 2-3 weeks). In addition, most of the LPPL studies of
bubbles previously performed (see (Sornette, 2003) and references
therein) have shown that the maximum of a bubble is not
immediately followed by a crash but that a rather complex behavior
(lasting a few days to a few weeks) develops before the crash
occurs. Here, we propose a simple model of the interplay between
prevailing opinions and personal preference of investors to buy or
sell shares which provides a useful classification of evolutionary
patterns of stock price dynamics including the so-called
``critical zone.''

\section{A short history of Ahold}

Albert Heijn started in 1887 with a small grocery store in
Zaandam, just north of Amsterdam. The Netherlands-based holding
company Ahold was listed on the AEX, the Amsterdam Stock Exchange,
in 1948 and opened its first self-service supermarket in Rotterdam
in 1955. The operating company of Albert Heijn became the
supermarket leader in the Netherlands and a household word for
quality and value-for-money. Shortly after the Queen awarded the
company the designation ``Royal,'' the last Heijn retired in 1989
as CEO from the company. Under his leadership, Royal Ahold had
started already in the 1980s an expansion program in the USA.
Ahold is the largest food provider in the Netherlands and one of
the largest in the United States. At the end of 2002, Ahold
operated some 5600 stores and employed approximately 280,000
people. Its major operations are in Europe and the United States,
but is has also expanded into retail operations in Latin America
and Asia as well. In the latter regions the company is now in the
process of divesting all its operations.

In 1993, Cees van der Hoeven became the new CEO. He had the at
first widely applauded ambition to make Ahold one of the largest
food providers in the world. At that time, Ahold's sales were
about 10 billion euro's. When he was forced to resign ten years
later, in 2003, the sales amounted to almost 63 billion euro's, a
six-fold increase. But then, the company almost collapsed under
its debt burden of 13 billion euro's, fraud and mismanagement. It
is generally assumed that the acquisition in 2000 of U.S.
Foodservice, a non-core activity,  which in itself can be viewed
as a strategic mistake, was the start of a host of problems
including the bookkeeping fraud at this same company. On 24
February 2003, the Ahold bubble collapsed and, while the stock
price on the AEX had been coming down under the threat of profit
warnings during the whole of 2002 from a high of 35 euro in 2001
to about 10 euro at the end of the year, it fell on that day with
a bang of $63\%$ down to 3.6 euro. The crisis was deep and
prolonged for customers, investors and employees alike.

\begin{figure}[htb]
\begin{center}
\includegraphics[width=7cm]{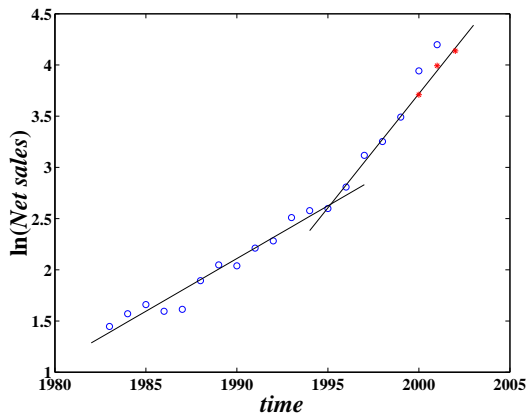}
\end{center}
\caption{ Annual ln(sales) (billion euro) of Ahold fitted with two
linear regression lines indicating an acceleration of the
exponential growth rate at the intersection of the lines in 1996.
The lower line was fitted to the data from 1983 to 1995 and
corresponds to $\ln ({sale})= 0.103 (t-1900)-7.16$. The upper line
is a fit of the data from 1995 to 1999 and corresponds to $\ln
({sales})= 0.223 (t-1900)-18.58$. The sales data of 2000 and 2001
indicate another growth acceleration. The corrected 2000 and 2001
sales data and the 2002 sales, all reported in October 2003 by
Ahold, are shown to coincide with the upper regression line. }
\label{Fig2exp}
\end{figure}

Inspection of a simple diagram (Figure \ref{Fig2exp}), which shows
the natural logarithm of the annual sales of Ahold from 1983
onwards, ln(sales), is already quite revealing. A straight line in
this plot indicates exponential growth. The comparison of the two
lines in Figure \ref{Fig2exp} is suggestive of an acceleration of
the growth rate of net sales in 1996. From the regression lines,
it can be seen that the average growth rate has more than doubled
from the first (1983-1995) to the second period (1995-1999). The
beginning of the second period corresponds approximately to the
time when Ahold started to consolidate the results of their latest
acquisition, the US supermarket chain Stop \& Shop. Also in 1996,
Royal Ahold became, in more than one way, an ``American company''
rather than a Dutch one because, for the first time, Ahold's sales
in the USA exceeded 50\%  of total sales bypassing the sales in
The Netherlands (41\%), which in the previous year still
represented the largest share of 48 \%. Furthermore, in the years
2000 and 2001, the growth appeared to accelerate again (The
corrections on the reported sales made in October 2003 indicate
that the sales figures were somewhat ``blown up'' since they
better fit the extension of the 1995-1999 regression line). Though
revealing, this simple description in terms of two, three or more
periods with different growth rates falls short of capturing
adequately the behavior of the Ahold sales as we discuss in the
next section.

\section{``Super-exponential'' growth of Ahold's sales, earnings and stock market prices}

\subsection{Super-exponential growth due to positive feedback}

Figure \ref{Fig2exp} actually shows a super-exponential growth in
sales. Such super-exponential behavior can be explained by the
concept of ``positive feedbacks,'' that is, conditioned on the
observation that the sales or market have recently moved up
(respectively, down), this makes them more probable to keep them
moving up (respectively, down), so that a large cumulative move
ensues. The concept of ``positive feedbacks'' has a long history
in economics and is related to the idea of ``increasing returns,''
which says that goods become cheaper the more of them are produced
(and the closely related idea that some products, like fax
machines, become more useful the more people use them). Positive
feedback is the opposite of negative feedback which is well-known
in population dynamics: the larger the population of rabbits in a
valley, the less grass there is per rabbit. If the population
grows too much, the rabbits will eventually starve, slowing down
their reproduction rate, which thus reduces their population at a
later time. Thus negative feedback means that the higher the
population, the slower the growth rate, leading to a spontaneous
regulation of the population size. In finance, value investing for
instance leads to negative feedbacks: if the observed price of a
company is larger than the estimated fundamental value, a value
investor will tend to sell, expecting the price to converge back
in the future to the true value. But by selling, he also tends to
push the price down, the very expected move that led to the
investment decision in the first place.

Positive feedback is the opposite phenomenon. In the context of
population dynamics, positive feedbacks have characterized the
evolution of human population growth over most of the last two
thousand years (see (Johansen and Sornette, 2001) and references
therein). The difference between humans and rabbits in this
context is that humans modify the carrying capacity of the planet
by various agriculture and technological innovations: the larger
the human population, the more probable the occurrence of
improvements and the exploitation of new habitats and resources,
etc., leading to a positive feedback of population on food
availability, providing a positive feedback on the growth rate and
so on. The consequence is that, until recently, the growth rate of
the human population has grown itself, leading to a
super-exponential growth of the population (one could say that
Malthus was an optimist!). For stock markets, when positive
feedbacks dominate, the higher the price or the price return in
the recent past, the higher will be the price growth in the
future. Apart from technical mechanisms for positive feedback such
as derivative hedging and portfolio insurance strategies,
behavioral traits of investors such as imitation and herding akin
to self-fulfilling prophecies play an important role (see Chap. 4
of (Sornette, 2003) for a detailed discussion and references
therein). Positive feedbacks, when unchecked, can produce runaways
until the deviation from equilibrium is so large that the growth
becomes unstable so that any disturbance/news may lead to ruptures
or crashes (Sornette, 2003). See (Sornette et al., 2003) for
examples of run-away hyperinflation due to positive feedbacks.

To illustrate the concept of positive feedback in a mathematical
form, consider the variable $X$ (net sales, or earnings, or
logarithm of prices). The familiar picture of a healthy growth of
$X$ is related to an exponential
\begin{equation}
  dX/dt = rX~,
  \label{Eq:dXdt}
\end{equation}
where $r$ is the instantaneous growth rate. For $r=r_0$ constant, $X$
grows exponentially with the constant growth rate $r$. We
interpret $r_0$ as the natural and normal growth rate for instance
associated with population growth and gains of productivity. Since
gains of productivity are usually small (Fair, 2002; van
Biesebroeck, 2004) and population growth is very small for a
developed country, $r_0$ is small for a company that has already a
substantial part of the local market as is the case of Ahold.

However, top managers are usually not satisfied with such small
growth rates and attempt to increase growth for example by making
acquisitions in foreign countries, which is partly due to the fact
that they can perceive more positive benefits from high growth
rates (Erickson,~Hanlon~\&~Maydew,~2004). Then $r$ itself becomes
a function of time. There are several ways to describe this
phenomenon. For instance, one can argue that $r$ increases with
$X$ due to expansion through acquisitions (like in Ahold's case)
according to \be r = r_0 + a X^k, \label{Eq:r} \ee where $k>0$ and
$a$ a positive constant. This expression embodies the concept of
positive feedback of the sales on its growth rate. Putting this
dependence (\ref{Eq:r}) in (\ref{Eq:dXdt}) yields
\begin{equation}
 dX/dt = r_0 X + (X/X_0)^{k+1}~,
 \label{Eq:gX}
\end{equation}
where $X_0=a^{-1/(k+1)}$ is a constant. Since $k>0$, the positive
feedback in (\ref{Eq:gX}) is initially small for small $X<X_0$ and
then accelerates progressively as $X$ reaches and overpasses the
characteristic value $X_0$. For $X>X_0$, the solution of
(\ref{Eq:gX}) tends to the asymptotic power law \be X(t) \propto
\frac{1}{(t_c -t)^{1/k}}~, \label{mgmel} \ee where $t_c$ is a
constant of integration. The solution (\ref{Eq:gX}) leads to what
is called in mathematical terms a ``movable finite-time
singularity'': ``singularity'' because of the divergence in finite
time as $t \to t_c$ and ``movable'' because the critical time
$t_c$ is not fixed a priori by the structure of the feedback and
of the dynamics but is sensitively dependent upon (and fixed by)
the initial conditions.

The acceleration of growth can also take the form of a feedback
of the velocity of change of the sales on the growth rate, such that
\be
r = b (dX/dt)^q
\label{nrfl}
\ee
with $0 \leq q < 1$. In words, the larger the slope of variation of the sales,
the larger the number of acquisitions and therefore the larger the
growth rate and so on.  Placing (\ref{nrfl}) in (\ref{Eq:dXdt})
leads to an equation of the form (\ref{Eq:gX}) (without the linear term
in the r.h.s.) with $1<k+1$ replaced by $1<1/(1-q)$. This leads to
$X(t) \propto 1/(t_c -t)^{1-q \over q}$. Note that the limit $q \to 0$
recovers the standard constant growth rate, with an exponential growth
formally retrieved as the limit of a power law with an infinite exponent.
These two models (\ref{Eq:r}) and (\ref{nrfl})
are end-members of more general feedback mechanisms
that can combine both effects.

The important point is that a rather broad class of positive
feedback mechanisms give rise to a power law acceleration, which
can be written generally as
\begin{equation}
 X(t) = A + B (t_c-t)^{m}~,
  \label{Eq:PL}
\end{equation}
where $m= -\frac{1}{k}= - (1-q)/q <0$ and $A$ and $B>0$ are two constants.
In particular, the constant $A$ has been added to represent the first correction
to the leading power law behavior valid for $t$ close to $t_c$.
Note that model (\ref{Eq:PL}) is much more parsimonious than a
model such as suggested in Figure \ref{Fig2exp} consisting in
several distinct regimes of exponential growth.

The form of the solution (\ref{Eq:PL}) illustrates the concept
that super-exponential growth hides an inherent danger. Though
theoretically, it may appear to go to infinity, what everybody
knows is that in practice it can not continue indefinitely. The
finite-time singular structure of (\ref{Eq:PL}) is actually the
mathematical translation of a change of regime, a cross-over from
an unsustainable pace to a crash or maybe a smooth landing of the
accelerating bubble regime (Johansen et al., 1999; 2000).

\subsection{Super-exponential growth in net sales and earnings}

\begin{figure}
 \centering
 \includegraphics[width=7cm]{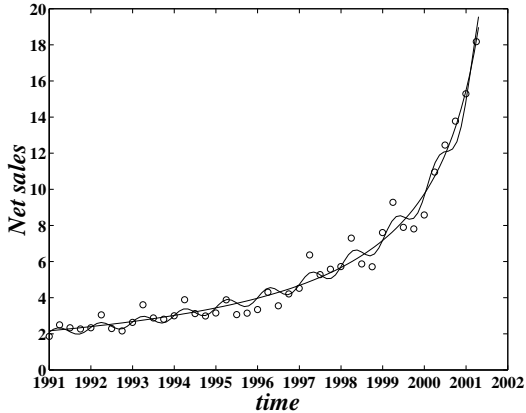}
\caption{Super-exponential growth of the quarterly net sales
(Billion Euro). The lines are respectively the fits to a pure
power law model (\ref{Eq:PL}) and to a periodically oscillatory
power law model (\ref{Eq:PLcos}). The power-law fit gives
$t_c={\rm{2002/05/20}}$, $m=-0.79$, $A=-0.975$, and $B=2185$ with
a r.m.s of fit residual equal to $0.609$. The oscillatory
power-law fit gives $t_c={\rm{2002/04/29}}$, $m=-0.65$, $f_s =
1.0006/365$, $\phi=4.99$, $A=-2.24$, $B=983.5$, and $C=-54.4$ with
a r.m.s. of fit residuals equal to $0.5186$.} \label{Fig:41}
\end{figure}

\begin{figure}
 \centering
 \includegraphics[width=7cm]{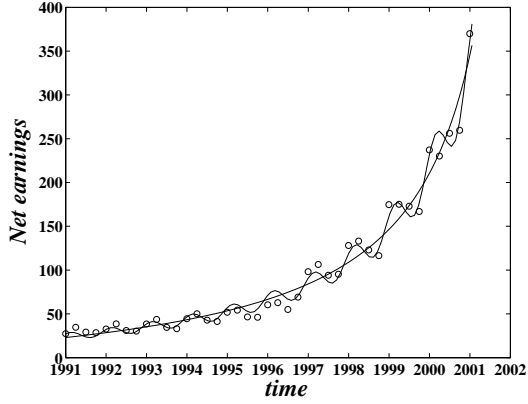}
\caption{Super-exponential growth of the quarterly net earnings
(Million Euro). The lines are respectively the fits to a pure
power law model (\ref{Eq:PL}) and to a periodically oscillatory
power law model (\ref{Eq:PLcos}). The power-law fit gives
$t_c={\rm{2002/09/28}}$, $m=-0.95$, $A=-39.8$, and $B=17882$ with
a r.m.s of the fit residuals equal to $13.7$. The oscillatory
power-law fit gives $t_c={\rm{2004/08/02}}$, $m=-1.77$, $f_s =
0.9989/365$, $\phi=1.75$, $A=-8.43$, $B=1.1\times 10^8$, and
$C=1.3\times 10^7$ with a r.m.s. of fit residuals equal to
$8.26$.} \label{Fig:42}
\end{figure}

Figures \ref{Fig:41} and \ref{Fig:42} expand on Figure
\ref{Fig2exp} and show respectively the quarterly net sales (from
1991 to 2001) and net earnings (from 1991 to 2001) of Royal Ahold.
The fits by the power law (\ref{Eq:PL}) of the quarterly net sales
(Billion Euro), net earnings (Million Euro) and logarithm of stock
prices of Ahold from 1991 to their historical highs at $t_{\max}$
are shown in figures \ref{Fig:41} and \ref{Fig:42}. For the
quarterly net sales, $t_{\max}={\rm{2001/03/31}}$. For the net
earnings, $t_{\max}={\rm{2000/12/31}}$. These three time series
exhibit a clear super-exponential growth. This occurred at a time
when the company was aggressively increasing its share of the
world market, transforming itself into an ever faster growing
entity.

Inspection of the figures  \ref{Fig:41} and \ref{Fig:42}
shows that the last quarter is always by
far the highest sales and earnings of the year. We can take into
account this yearly periodicity by the improved model
\begin{equation}
 X(t) = A + B (t_c-t)^m + C (t_c-t)^m \cos(2 \pi f_s t + \phi)~,
 \label{Eq:PLcos}
\end{equation}
where $f_s$ is the fundamental frequency of the seasonal cycle.
The theoretical value of $1/365$ (for $t$ in units of day) of a
yearly cycle is confirmed by fitting. The fits of the sales and
earnings with (\ref{Eq:PLcos}) are also shown in Figures
\ref{Fig:41} and \ref{Fig:42}. A standard Wilks test (Rao, 1966)
for the statistical significance of the added term $C (t_c-t)^m
\cos(2 \pi f_s t + \phi)$ gives a log-likelihood ratio between the
models (\ref{Eq:PL}) and (\ref{Eq:PLcos}) equal to $13.5$, so that
the probability that the added explanatory power provided by
(\ref{Eq:PLcos}) over (\ref{Eq:PL}) results from chance is
$0.12\%$. The statistical significance of model (\ref{Eq:PLcos})
is thus established at the $99.9\%$ confidence level. Similarly,
the corresponding log-likelihood ratio between the models
(\ref{Eq:PL}) and (\ref{Eq:PLcos}) for the  net earnings is equal
to $41.5$, so that the probability that the explanatory power of
(\ref{Eq:PLcos}) results from chance is essentially zero.

\subsection{Bubble and post-bubble regimes in stock prices}

\begin{figure}[htb]
\begin{center}
\includegraphics[width=7cm]{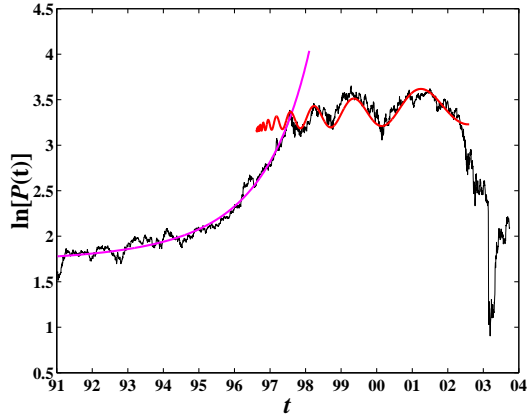}
\end{center}
\caption{Ahold stock price as a function of time. The vertical
axis is in logarithmic scale while the horizontal axis uses a
linear scale. The power-law fit to the data from $\rm{1991/01/10}$
to $t_{\rm{max}}=\rm{1997/07/23}$ gives $t_c={\rm{2004/06/19}}$,
$m=-4.62$, $A=1.71$, $B=8.11\times10^{15}$ with a r.m.s of the fit
residuals equal to $0.062$. The log-periodic power-law fit to the
data from $\rm{1997/04/09}$ to $\rm{2002/08/09}$ gives
$t_c={\rm{1996/07/24}}$, $\alpha=0.48$, $\omega=12.31$,
$\phi=3.61$, $A=3.14$, $B= 0.0078$, $C= 0.0055$, and the r.m.s. of
the fit residuals is $\chi=0.103$.} \label{Fig:33}
\end{figure}

For the stock prices of Royal Ahold, the super-exponential bubble
culminated on $t_{\max}={\rm{1997/07/23}}$. Figure \ref{Fig:33}
shows the super-exponential growth of the logarithm of Ahold stock
prices.

The post-bubble regime since the local high on $\rm{1997/07/23}$
has also been fitted with a log-periodic post-bubble power law
(Johansen and Sornette, 1999; Sornette, 2003)
\begin{equation}
 \ln \left[P(t)\right] = A + B (t-t_c)^m + C (t-t_c)^m \cos\left[\omega \ln(t-t_c) + \phi\right]~,
 \label{Eq:LPPL}
\end{equation}
showing an oscillatory pattern of increasing amplitude. The fit is
also illustrated in Fig.~\ref{Fig:33}. It is striking to note that
$t_c$ is also located in 1996, the point in time where according
to the simpler analysis in Fig. 1 the growth acceleration took
place. This reinforces the conclusion that 1996 is the year where
a change of regime took place and the seed of the 2002/03 crisis
was planted initiating a ``critical zone'' of destabilization.
This pattern will be discussed more thoroughly later in the paper.

These analyses suggest that the crisis was not due to a proximate
event but was seeded a long time ago when the policy of
accelerating growth started to be implemented in 1996. Sornette
(2003) has synthesized a large body of the academic literature
showing that the general reason for a crash or crisis is rarely
due to proximate causes but results from the progressive
maturation and emergence of an unstable phase. When the unstable
phase is ripe, basically any perturbation may trigger the crisis.
This concept allows to understand why it is often so difficult to
explain or understand the origin of the crisis or why so many
different explanations are sometimes advanced, when one insists in
searching for a proximate origin. The fundamental origin based on
a growing instability could be called ``emergent'' and
``systemic'' in the jargon of the theory of complex systems. For
these, one needs to develop a description of the collective
bottom-up behavior rather than attempting to find a smoking gun.

An important question concerns the predictive power of such fits.
We should stress that no precise timing can be claimed here, for
instance by using the fitted $t_c$ as a prediction for an
impending crash. Our analysis performed jointly on the sales,
earnings and prices can only be used to point to the approach of a
change of regime which is indicated by the critical times. Results
from fitting procedures using truncated time series on general
markets like the S\&P 500 have shown that, as a rule of thumb, the
critical time $t_c$ can be ``rather robust to approximately one
year prior to a crash'' (Sornette, 2003, p.330). The determination
of the precise timing seems more complex for individual companies
than for the overall market bubbles ending in crashes documented
in (Sornette, 2003), probably due to several factors: the
interference with the global market mood, the coupling with the
policy of acquisitions and the manipulations of the corporate
performance data as well as other idiosyncratic factors.
Nonetheless it is striking to note that the above rule of thumb
also appears to apply to the prediction of the critical times in
2002 from the sales data (Fig. 2) when fitted by (\ref{Eq:PL}) and
(\ref{Eq:PLcos}). Using a truncated fitting procedure this result
of a $t_c$ in 2002 appears to be rather robust. Although
(\ref{Eq:PL}) also predicts a $t_c$ in 2002 for the earnings data,
this is not a robust result. This comes as no surprise since
quarterly earnings data are much more subject to manipulation
showing no consistent results.

So, whereas the super-exponential growth of sales and earnings can
not be expected to give precise warnings of an impending crash
within at most one year, combining these data with the stock
market price evolution opens new possibilities for advanced
warnings. As shown in Fig. \ref{Fig:33}, 4-5 years before the
crash in 2002 the stock price has changed from super-exponential
growth to a different oscillatory regime. It is also clear that
the transition from the super-exponential growth to the crash did
not proceed directly but through an intermediate regime
punctuating the evolution. The model we discuss below suggests an
interpretation in terms of a variation with time of the opinion
formation process among investors after the peak of the bubble,
during the critical zone and then during the crash. The important
point of our analysis is to focus on medium-term evolutionary
patterns rather than on short-term vagaries. This strategy leads
to the possibility of recognizing years before the actual crash
occurred that a company, while its sales and earnings are racing
upward, is really moving through a critical zone which, if allowed
to continue, makes a crash inevitable.

\section{Weidlich's model of opinion formation applied to the stock market}

\subsection{Motivation}

To summarize, we have argued that the evidence of positive
feedbacks on the sales, earnings and stock market price of Ahold
can be characterized by the occurrence of an apparent finite-time
singularity. This singularity must be understood as announcing a
change of regime. Usually, this change of regime takes the form of
a break in the super-exponential growth which transitions into
what we propose to define as a ``critical zone.'' The maturation
of the critical zone may last weeks, months to years (rarely), and
is usually followed by a catastrophic crash. Our motivation for
introducing the concept of a ``critical zone'' is based on the
following.
\begin{itemize}
\item  Previous related works
have found that the crash does not immediately follow the top of the stock
market bubble but there is a transitional period (Sornette, 2003 and references therein).
\item  Models of rational expectation (RE)  bubbles incorporating the mechanism
of positive feedbacks (Johansen et al., 1999; 2000; Sornette and Andersen, 2002)
predict a very large
crash hazard rate in the vicinity of the end of the bubble; in such models,
there is an effective ``critical zone'' characterized by a very large
risk of a crash, without the need for the price to continue to accelerate.
Since RE models of bubbles are only providing coarse-grained descriptions,
it would be interesting to develop finer models of this regime.
\item There are general arguments to expect that a finite-time singularity
is rounded by finite-size effects
(Cardy, 1988) and by the triggering of
negative feedback mechanisms that were sub-dominant but come progressively
into play as the dynamics approach the singularity.
This rounding may be part of the ``critical zone.''
\end{itemize}
We now turn to the development of a model of the
``critical zone'' and other patterns as well, based on Weidlich's
approach to social models using concepts from synergetics, which
we will use to describe and interpret the evolution of Ahold stock
price.

\subsection{Background of the Weidlich model}

Synergetics was originally developed by the German physicist Hermann
Haken (1983) to study the behavior of complex systems of any kind.
Composed of many interacting parts, these systems, whether physical,
biological or social, are known to be able to spontaneously form
analogous spatial, temporal or functional structures or patterns through
self-organization. Since its focus is particularly on what happens in
those situations where complex system change their behavior
qualitatively, Haken (2000) also considered synergetics as a theory of
emergence of new qualities at a macroscopic level. The mechanisms
underlying the observed processes at non-equilibrium phase transitions
and bifurcations are studied with concepts of instability, order
parameters and slaving. The slaving principle yields the important
insight that close to instability points complex systems are governed by
a low-dimensional, though noisy, dynamics (Haken, 2000).

Haken's colleagues, Weidlich and Haag (1983), focused synergetics on
social science applications such as the dynamics of the political
opinion formation process and, in economics, on the non-equilibrium theory of
investment behavior, also known as `Schumpeter's clock.' Studying the
dynamics of two types of investment projects, expansionary and
rationalizing through a model that formally resembles the opinion
formation model, Weidlich and Haag (1983) were able to faithfully
reproduce the evolutionary pattern of industrial strategic investment in
Germany between 1956 and 1978. Under the title Sociodynamics, Weidlich
(2000) published an elaborated version of the formal opinion formation
model and included further examples of patterns of political phase
transition and destabilization which are accompanied by unpredictable
critical fluctuations where `anything can happen.' The formal model,
though relatively simple, is very powerful and has also been applied to
the phenomenon of lock-in of a dominant technology in a situation of
competing technologies (Arthur, 1994) and the inverse phenomenon of the
lock-out of a dominant technology or product through disruptive
innovations (Broekstra, 2002).

We will here closely follow Weidlich's (2000) extended formal model of
opinion formation and political phase transitions which has a strong
resemblance to the dynamics of investors' opinions or decisions as to
`buying' or `selling' shares in the stock market. Particularly, the
discontinuity that occurs in the transition from a liberal democratic to
a totalitarian political system or vice versa (the lib-tot phase
transition) is argued to have a close analogy to the development of
bubbles and crashes in the financial markets. Political phenomena like
public pressure to conform to the ruling ideology on the one hand and
dissident behavior on the other can equally be transferred to the
economic domain where the combination of imitation pressures and herding
on the hand and idiosyncratic or contrarian behavior on the other may be
responsible for potentially unstable situations in stock markets
(Sornette, 2003).

Simple models of dynamic behavior of investors in the stock market
consider investors to be in one of only two possible states, such
as `buy' or `sell,' `optimistic' or `pessimistic,' and `bullish'
or `bearish.' In addition, several models have considered at least
two types of investors such as `trend followers' and `value
investors' (e.g. Farmer, 1998; Lux and Marchesi, 1999; Levy et
al., 2000; Ide and Sornette, 2002). Here, we will not assume from
the outset that there are two rigid types of investors who
basically account for positive and negative feedback processes,
but we will introduce parameters for a population of investors
holding `buy' or `sell' opinions. These control parameters can
co-evolve with the order parameters to determine the degree of
public conformity with the prevailing opinion, which influences
the degree of imitation and herding, and the personal inclination
to approve or disapprove of the prevailing opinion, which governs
the degree of affirmation or dissidence with it. This allows us to
investigate all possible patterns of evolution of share prices
and, particularly, the complete patterns of destabilization of the
price trajectories other than bubbles and crashes alone. We will
show that, because of the flexible psychology of investors, the
dynamics of one of the parameter of the model controlling trends
will play an important if not decisive role. Describing the
dynamics of such control parameters in an endogenous manner
massively complicates the structure of the model. Furthermore, it
will not be simple to find plausible systems of equations of
motion for them. Such equations would also have to depend on the
global economical and political situation, i.e., not only on the
endogenous variables described below following Weidlich (2000).
Therefore, we choose the simpler but efficient method (also used
in the Schumpeter clock model (Weidlich and Haag, 1983)) to
distinguish different time periods, each with plausibly chosen
trends to evaluate the stock price evolution of Ahold. This
requires a careful interpretation based on detailed information on
the historical development of the Ahold drama.

Naturally, a lot of attention has focused on the development of
bubbles and crashes, but hardly to what happens in the time period
right after a super-exponential growth spurts, when the bubble has
run out of steam, and before a crash may occur. This intermediate
stage preceding a crash turns out to be a period of increasing
destabilization, which may take weeks, months or years.
Furthermore, a lot more attention has been given to stock market
indices than to the fate of individual companies.

\subsection{The evolution equations of majority and personal opinions}

We first consider the evolution of the so-called ``investors'
configuration'', $\{n_B(t), n_S(t)\}$, where it is assumed that at
a given point in time the sum of  the time-varying numbers of
buyers $n_B(t)$ and sellers $n_S(t)$ is a constant and equal to
$2N$. The investors' configuration can then be defined by the
normalized variable $y(t)$ defined by \be y(t) =
\frac{n_B(t)-n_S(t)}{2N}~, ~~~{\rm where}~~ -1 \leq y(t) \leq 1~.
\label{2} \ee When $y(t) = 0$, the two types of investors holding
opinions `buy' or `sell' are in balance: $n_B(t) = n_S(t)$. Supply
and demand are equal and the share price remains constant. If the
number of buyers is larger (smaller) than the number of sellers,
$y(t) > 0$ (resp. $y(t)< 0$), then demand exceeds (resp. is lower
than) supply and the price will go up (resp. down).

Investors may hold a personal inclination or preference concerning
buying or selling, which may approve or disapprove of the
prevailing opinion. For example, during a bubble formation when
the psychology of herding and imitation takes over, traders may be
forced to conform to pressures from their clients to buy even when
they privately disapprove. This effect is taken into account by a
personal preference variable $m_i(t)$ where $i$ is either the
`buy' or `sell' state, such that a value $m_i(t) > 0$ describes an
individual's personal preference for the $i$-th state, a value
$m_i(t)= 0$ describes inner neutrality to the state $i$; a value
$m_i(t)< 0$  describes personal disapproval of the $i$-th state.
Assuming that $m_B = - m_S =  m(t)$ varies between limits, $-M
\leq m(t) \leq M$, a normalized variable (the personal preference
index) is introduced
\be
x(t) = m(t)/M~, ~~~~{\rm where}~~  -1
\leq x(t) \leq 1~.
\label{nmmlmdf}
\ee
If $x( t) = 0$, personal
preferences towards buying and selling are neutral; if $x( t) > 0$
(resp. $x(t) < 0$), personal inclinations are to approve of the
prevailing opinion to buy (sell) and disapprove of selling
(buying). The more positive or negative $x(t)$ is, the stronger
the growth of the personal preference towards one or the other
position.

The quantities that govern the dynamics of investment decision
making are the individual transition probabilities per unit time
period of a `buy' to a `sell' state or opinion and vice versa, and
the individual transition rates of the personal preference towards
higher (lower) values of approval or disapproval. An individual
changes its opinion randomly depending on the aggregated average
opinion found in the population and on his own opinion. Thus,
individuals do not interact directly at the micro level, but
indirectly via the macro level (see for instance Helbing, 1992 for
a generalization including pair interactions). In Weidlich's
(1983, 2000) theory, an ensemble of configurations is considered
where each is a homogeneous population of $2N$ investors. The
probability distribution over these configurations $p(x,y;t)$ is
then defined as the probability that one sample configuration has
the configuration $\{x,y\}$ at time $t$. A nonlinear differential
equation is then defined, the master equation, which describes the
changes of the probability distribution over time. For further
details the reader is referred to Weidlich (2000; see also
Broekstra, 2002).

Approximate solutions to the probabilistic master equation can be
obtained by considering the quasi-meanvalue equations for the mean
evolution of the order parameters $x$ and $y$. This is the
approach followed here. However, for a fuller understanding of the
origin of the quasi-meanvalue equations to follow, it is noted
that Weidlich (2000) proved  without loss of generality that the
transition rates must be of the form $m \exp (u ({\rm final})-u
({\rm initial}))$, where $m$ is a mobility factor and $u(initial)$
and $u(final)$ are utility functions depending on the system
variables and personal state before and after the transition. The
actual form of the utility function still has to be determined
and, for reasons of simplicity, was based by Weidlich on a linear
dependence on the number of investors holding the `buy' or `sell'
opinion (Weidlich, 2004, private communication). The quasi-mean
value equations consist of a system of two nonlinear differential
equations for the normalized personal preference $x(t)$ and
investors configuration $y(t)$ (Weidlich, 2000): \ba \frac{d
y}{dt} &=& \sinh (\kappa y + \gamma
x) - y ~\cosh (\kappa y + \gamma x)~, \label{y1eq} \\
\frac{d x}{dt} &=& \mu \left[ \sinh (\beta y) - x ~\cosh (\beta y ) \right]~.
\label{x1eq}
\ea
To simplify the notations, $x$ and
$y$ are now to be read as mean values and the time $t$ is also a
scaled variable. Equations (\ref{y1eq}) and (\ref{x1eq}) have four
parameters $\kappa, \gamma, \beta$ and $\mu$, which describe
behavioral trends and express the investors' psychology. We will
treat them as constant in each relevant regime.

$\kappa$ is a parameter coupling individuals to the majority. We
refer to it as the coupling or conformity parameter. If positive,
it strengthens the investors' personal readiness to conform to the
majority opinion. If it is high, then the influence of a `buy'
majority on an individual transition from `sell' to `buy' is high.
Inspection of Eq.(\ref{y1eq}) shows that if $y > 0$, the existing
majority opinion to `buy' further favors transitions from `sell'
to `buy', and similarly discourages the inverse transitions. As we
will see, if $\kappa$ attains a critical value, a herding process
starts which can lead to a bubble. Conversely, if $\kappa> 0$ and
$y < 0$, a majority opinion to `sell' will put pressure on
individual investors to conform by changing their position from
`buy' to `sell'. As Eq.(\ref{y1eq}) shows, the evolution of the
majority also depends on the investors' personal preference $x$.
Clearly, with $\gamma >0$ and personal preference $x$ is positive,
these effects will be reinforced. If however $x < 0$ and investors
tend to personally disapprove of the majority `buy' position, this
will diminish the effect of the majority pressure to conform. In
fact, this idiosyncratic or contrarian inclination, which Weidlich
(2000) called a `dissidence propensity' in the case of a political
opinion formation process, may eventually lead to the
destabilization of the system.

The inclination parameter $\beta$ governs the evolution of the personal
preference. If positive, it measures an approving tendency and, as Eq.(\ref{x1eq})
shows, when the majority $y > 0$ favors the `buy' decision, the evolution
of a positive $x$ will be reinforced stabilizing the majority opinion.
However, if $\beta$ is sufficiently negative, that is, attains a critical
value, the dissident propensity may lead to a strong tendency to
personally disapprove of the existing situation, eventually leading to a
negative $x$, which in turn may lead to a destabilization of the majority
position.

The preference influence parameter $\gamma$ determines the strength of the
influence of the personal preference $x$ on the evolution of the majority
opinion, and $\mu$ is a parameter which determines the speed of evolution of
the personal preference relative the speed of evolution of the majority opinion $y$.
In other words, $\mu$ is the ratio of the characteristic time scales of the
dynamics of the majority opinion to the dynamics of personal preference. A small
(large) $\mu$ corresponds to a slow (fast) adjustment of personal preference
to majority opinion.


We complement this system (\ref{y1eq}) and (\ref{x1eq}) by an
equation for the price. We make the simplified assumption that the
mean individual order sizes $O_B$ for buyers and $O_S$ sellers are
equal and constant, such that $O_B = - O_S = O$. Then, the net
order size of the population is $\Omega(t) = n_B O_B + n_S O_S = O
(n_B - n_S)  = 2N O y(t)$, using (\ref{2}). Dividing both sides by
the size $2N$ of the population, we obtain the scaled linear
equation for the net order size per agent \be \omega(t) = \alpha
y(t) \label{9} \ee where $\alpha$ is a positive constant. If
supply and demand balance each other, $y(t) = 0$, and the net
order size equals zero. If buyers prevail over sellers (or vice
versa), $y(t) > 0$ ($y(t) < 0$), the net order size is positive
(negative). To make the theory simple, we assume following many
previous workers that the market impact function which relates the
logarithm of the stock price $P$ to the net order size $\Omega$
reads \be \frac{d\ln P(t)}{dt} =  \Omega(t)/L~, \label{10} \ee
where $L$ is some `market depth' assumed to be constant. Scaling
this equation, and henceforth for reasons of convenience using the
notation $p(t)$ for $\ln P(t)$ for the mean value of the stock
price, in the numerical integration we used the simple equation
for the logarithmic stock price evolution \be \frac{d p(t)}{dt} =
a y(t)~, \label{11} \ee where $a$ is a positive constant. This
expression (\ref{11}) implies that the logarithm $p(t)$ of the
price is proportional to the integral of the majority opinion
order parameter $y(t)$ and reciprocally $y(t)$ is proportional to
the market return $dp/dt$. The system we study below is thus made
of (\ref{y1eq}), (\ref{x1eq}) and (\ref{11}).

Using expression (\ref{11}) to eliminate $y(t)$ in the r.h.s. of
(\ref{y1eq}) and (\ref{x1eq}) offers an interesting interpretation
of the opinion formation process in stock markets: the proxy for
the majority opinion $y(t)$ is indeed the stock market return
$dp/dt$ which is the major factor influencing the dynamics of the
opinions. The following quip ``in the long run, markets are
weighing machines, but in the short term, they are voting
machines'' is often attributed to B. Graham and W. Buffet. This
quote exemplifies why the theory of Weidlich developed to model
opinion formation processes can indeed be a useful model of stock
market price evolution at short and intermediate times. At long
times, economic factors and fundamental valuations have to be
introduced, giving a competition between the ``voting'' and
``weighing'' properties of the stock market.

\subsection{Stability analysis}

The analysis of Eqs.(\ref{y1eq}) and (\ref{x1eq}) reveals that the
evolution equations exhibit five singular points, that is, points
where the left-hand side derivatives become equal to zero
(Weidlich and Haag, 1983; Weidlich 2000). Stability analysis of
the behavior of $x(t)$ and $y(t)$ in the vicinity of the singular
point $(x,y)= (0,0)$ is of particular interest because it deals
with the possibility of deviation from a balanced situation where
the number of buyers and sellers is equal and their personal
preference towards either decision is neutral. When adding noise
(stochastically), this stationary point $(0,0)$ corresponds to the
random walk situation and the question whether it is a stable or
unstable point becomes important to detect the emergence of a
runaway bubble or crash. Without going into further detail we will
just state the following stability conditions (see Weidlich,
2000).

The stationary point $(x,y)= (0,0)$ is stable if and only if \be
\kappa <  1 + \mu~~~~~  {\rm and}~~~~~  \beta < (1 -
\kappa)/\gamma~. \label{6} \ee The stationary state $(0,0)$ is
thus unstable if at least one of the two inequalities in (\ref{6})
is broken. These conditions (\ref{6}) express that a balanced
situation can continue to exist only if the pressure $\kappa$ to
conform to the majority opinion is relatively small, while
simultaneously the personal inclination $\beta$ to approve or
disapprove also remains small. Starting from a small degree of
conformity, if  $\kappa > 0$ starts to grow meaning that the
tendency to herd starts to increase, condition (\ref{6}) shows
that the value of $\beta$  has to become increasingly smaller and
even negative, indicating that the degree of dissidence among
investors needs to increase in order for the point $(0,0)$ to
remain stable. However, if $\kappa$ exceeds the critical threshold
value $1 + \mu$, whatever the amount of dissidence, $(0,0)$
becomes an unstable stationary point, and a runaway situation,
where the market locks in on either a bull market or a crash,  may
result. As (\ref{6}) also shows, if $\beta$ exceeds a critical
value even if $\kappa$ has not, $(0,0)$ becomes unstable. This is
reasonable to find that too much dissidence can equally
destabilize the neutral point.

Finally, a limit cycle solution or attractor of the mean value
equations exists if at least one of inequalities in (\ref{6}) is
broken and if in addition \be \beta < - \frac{(\mu + \kappa -
1)^2}{4 \mu \gamma} < 0 \label{8} \ee holds.

\subsection{Classification of evolution patterns}

The system of Eqs. (\ref{y1eq}) and (\ref{x1eq}) was numerically
integrated using a 4th order Runge-Kutta scheme, for a number of
different values of the parameters. In the numerical integrations
presented below, we have used a scaled time step of $0.01$
throughout. The numerical integrations have allowed us to classify
four basic patterns of evolution of the mean values of the
investors configuration $y(t)$, of the personal preference $x(t)$
and of the price $p(t)$, for different parameter values. First, to
attain some basic insights, we will keep the trend parameters
$\kappa$ and $\beta$ constant. However, during transitions, it is
more reasonable to assume that particularly the trend parameters
$\kappa$ as a measure of herding pressure and $\beta$ as a measure
of the degree of dissidence will co-evolve with the system
variables. This will be dealt with in the next section when we
discuss the evolution of the stock price value of Ahold.

 \subsubsection{Pattern A: $(0,0)$ stability and `random walk'}

\begin{figure}[htb]
\begin{center}
\includegraphics[width=7cm]{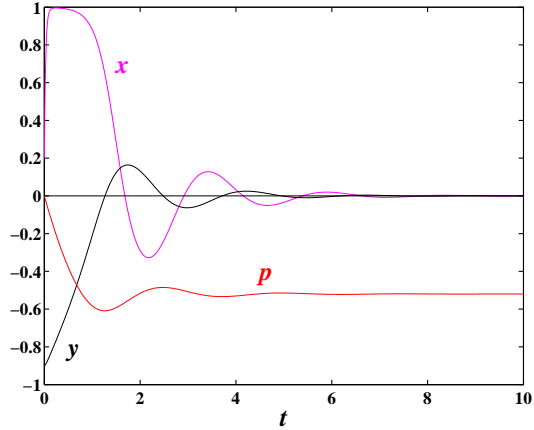}
\end{center}
\caption{Evolution Pattern A when $(x(t),y(t)) = (0,0)$ is a stable
 stationary state. This simulation
 corresponds to $\mu = 2$, $\gamma = 1$, $\kappa = 1.5$, $\beta = -4$, and $a = 1$.
 Since $\kappa = 1.5 < 1+\mu = 3$ and $\beta = -4 < (1-\kappa)/\gamma =
 -0.5$, according to condition (\ref{6}),  $(x,y) = (0,0)$ is a stable stationary
 state and any deviation from it will quickly return to that state.
For reasons of presentation, in this and the following
 figures, the logarithm of the stock price $p(t)$ starts at an arbitrary
 initial point $p(0)$.}
\label{Fig1}
\end{figure}

 This pattern is exemplified in Figure \ref{Fig1}, where a
 large negative initial value of $y(0)$, a sudden preponderance of sellers
 over buyers, is quickly restored to the balanced situation. Due to the
 large negative, disapproving $\beta$ used in this
 example, the personal preference index $x$ becomes
 highly positive, driving the out-of-balance situation quickly back to
 normal. Furthermore, since $\beta = -4 < -(\mu + \kappa -1)^2/(4\mu \gamma) = -0.8$,
 condition (\ref{8}) indicates that we may expect that the return will occur
 in an oscillatory manner. In contrast, for example with $\beta = -1$, the
 $y$-trajectory will approach the $(0,0)$ state somewhat more slowly, but
without oscillations.

\begin{figure}
\begin{center}
\setlength{\unitlength}{0.240900pt}
\ifx\plotpoint\undefined\newsavebox{\plotpoint}\fi
\begin{picture}(1949,1800)(200,0)
\font\gnuplot=cmr10 at 10pt
\gnuplot
\sbox{\plotpoint}{\rule[-0.200pt]{0.400pt}{0.400pt}}%
\put(220.0,113.0){\rule[-0.200pt]{4.818pt}{0.400pt}}
\put(198,113){\makebox(0,0)[r]{$200$}}
\put(1865.0,113.0){\rule[-0.200pt]{4.818pt}{0.400pt}}
\put(220.0,390.0){\rule[-0.200pt]{4.818pt}{0.400pt}}
\put(198,390){\makebox(0,0)[r]{$220$}}
\put(1865.0,390.0){\rule[-0.200pt]{4.818pt}{0.400pt}}
\put(220.0,668.0){\rule[-0.200pt]{4.818pt}{0.400pt}}
\put(198,668){\makebox(0,0)[r]{$240$}}
\put(1865.0,668.0){\rule[-0.200pt]{4.818pt}{0.400pt}}
\put(220.0,945.0){\rule[-0.200pt]{4.818pt}{0.400pt}}
\put(198,945){\makebox(0,0)[r]{$260$}}
\put(1865.0,945.0){\rule[-0.200pt]{4.818pt}{0.400pt}}
\put(220.0,1222.0){\rule[-0.200pt]{4.818pt}{0.400pt}}
\put(198,1222){\makebox(0,0)[r]{$280$}}
\put(1865.0,1222.0){\rule[-0.200pt]{4.818pt}{0.400pt}}
\put(220.0,1500.0){\rule[-0.200pt]{4.818pt}{0.400pt}}
\put(198,1500){\makebox(0,0)[r]{$300$}}
\put(1865.0,1500.0){\rule[-0.200pt]{4.818pt}{0.400pt}}
\put(220.0,1777.0){\rule[-0.200pt]{4.818pt}{0.400pt}}
\put(198,1777){\makebox(0,0)[r]{$320$}}
\put(1865.0,1777.0){\rule[-0.200pt]{4.818pt}{0.400pt}}
\put(333.0,113.0){\rule[-0.200pt]{0.400pt}{4.818pt}}
\put(333,68){\makebox(0,0){$87.8$}}
\put(333.0,1757.0){\rule[-0.200pt]{0.400pt}{4.818pt}}
\put(615.0,113.0){\rule[-0.200pt]{0.400pt}{4.818pt}}
\put(615,68){\makebox(0,0){$87.82$}}
\put(615.0,1757.0){\rule[-0.200pt]{0.400pt}{4.818pt}}
\put(897.0,113.0){\rule[-0.200pt]{0.400pt}{4.818pt}}
\put(897,68){\makebox(0,0){$87.84$}}
\put(897.0,1757.0){\rule[-0.200pt]{0.400pt}{4.818pt}}
\put(1179.0,113.0){\rule[-0.200pt]{0.400pt}{4.818pt}}
\put(1179,68){\makebox(0,0){$87.86$}}
\put(1179.0,1757.0){\rule[-0.200pt]{0.400pt}{4.818pt}}
\put(1462.0,113.0){\rule[-0.200pt]{0.400pt}{4.818pt}}
\put(1462,68){\makebox(0,0){$87.88$}}
\put(1462.0,1757.0){\rule[-0.200pt]{0.400pt}{4.818pt}}
\put(1744.0,113.0){\rule[-0.200pt]{0.400pt}{4.818pt}}
\put(1744,68){\makebox(0,0){$87.9$}}
\put(1744.0,1757.0){\rule[-0.200pt]{0.400pt}{4.818pt}}
\put(220.0,113.0){\rule[-0.200pt]{401.098pt}{0.400pt}}
\put(1885.0,113.0){\rule[-0.200pt]{0.400pt}{400.858pt}}
\put(220.0,1777.0){\rule[-0.200pt]{401.098pt}{0.400pt}}
\put(200,1900){\makebox(0,0){\large{S\&P $500$}}}
\put(1052,-23){\makebox(0,0){\large{time(year)}}}
\put(220.0,113.0){\rule[-0.200pt]{0.400pt}{400.858pt}}
\put(236,1475){\raisebox{-.8pt}{\makebox(0,0){$\Diamond$}}}
\put(275,1254){\raisebox{-.8pt}{\makebox(0,0){$\Diamond$}}}
\put(391,134){\raisebox{-.8pt}{\makebox(0,0){$\Diamond$}}}
\put(430,338){\raisebox{-.8pt}{\makebox(0,0){$\Diamond$}}}
\put(468,921){\raisebox{-.8pt}{\makebox(0,0){$\Diamond$}}}
\put(507,730){\raisebox{-.8pt}{\makebox(0,0){$\Diamond$}}}
\put(546,682){\raisebox{-.8pt}{\makebox(0,0){$\Diamond$}}}
\put(661,394){\raisebox{-.8pt}{\makebox(0,0){$\Diamond$}}}
\put(700,510){\raisebox{-.8pt}{\makebox(0,0){$\Diamond$}}}
\put(739,546){\raisebox{-.8pt}{\makebox(0,0){$\Diamond$}}}
\put(777,747){\raisebox{-.8pt}{\makebox(0,0){$\Diamond$}}}
\put(816,936){\raisebox{-.8pt}{\makebox(0,0){$\Diamond$}}}
\put(913,914){\raisebox{-.8pt}{\makebox(0,0){$\Diamond$}}}
\put(951,808){\raisebox{-.8pt}{\makebox(0,0){$\Diamond$}}}
\put(990,808){\raisebox{-.8pt}{\makebox(0,0){$\Diamond$}}}
\put(1029,881){\raisebox{-.8pt}{\makebox(0,0){$\Diamond$}}}
\put(1067,794){\raisebox{-.8pt}{\makebox(0,0){$\Diamond$}}}
\put(1183,745){\raisebox{-.8pt}{\makebox(0,0){$\Diamond$}}}
\put(1222,659){\raisebox{-.8pt}{\makebox(0,0){$\Diamond$}}}
\put(1261,698){\raisebox{-.8pt}{\makebox(0,0){$\Diamond$}}}
\put(1299,801){\raisebox{-.8pt}{\makebox(0,0){$\Diamond$}}}
\put(1338,773){\raisebox{-.8pt}{\makebox(0,0){$\Diamond$}}}
\put(1454,781){\raisebox{-.8pt}{\makebox(0,0){$\Diamond$}}}
\put(1493,705){\raisebox{-.8pt}{\makebox(0,0){$\Diamond$}}}
\put(1531,758){\raisebox{-.8pt}{\makebox(0,0){$\Diamond$}}}
\put(1570,651){\raisebox{-.8pt}{\makebox(0,0){$\Diamond$}}}
\put(1609,694){\raisebox{-.8pt}{\makebox(0,0){$\Diamond$}}}
\put(1725,725){\raisebox{-.8pt}{\makebox(0,0){$\Diamond$}}}
\put(1763,753){\raisebox{-.8pt}{\makebox(0,0){$\Diamond$}}}
\put(1802,727){\raisebox{-.8pt}{\makebox(0,0){$\Diamond$}}}
\sbox{\plotpoint}{\rule[-0.500pt]{1.000pt}{1.000pt}}%
\put(237,1469){\usebox{\plotpoint}}
\multiput(237,1469)(3.607,20.440){5}{\usebox{\plotpoint}}
\put(260.66,1538.84){\usebox{\plotpoint}}
\multiput(268,1526)(2.379,-20.619){7}{\usebox{\plotpoint}}
\multiput(283,1396)(1.600,-20.694){10}{\usebox{\plotpoint}}
\multiput(299,1189)(1.408,-20.708){12}{\usebox{\plotpoint}}
\multiput(316,939)(1.205,-20.721){12}{\usebox{\plotpoint}}
\multiput(331,681)(1.440,-20.705){11}{\usebox{\plotpoint}}
\multiput(347,451)(1.890,-20.669){9}{\usebox{\plotpoint}}
\multiput(363,276)(3.049,-20.530){5}{\usebox{\plotpoint}}
\put(391.30,160.14){\usebox{\plotpoint}}
\multiput(395,156)(5.197,20.094){3}{\usebox{\plotpoint}}
\multiput(410,214)(2.677,20.582){6}{\usebox{\plotpoint}}
\multiput(426,337)(1.991,20.660){8}{\usebox{\plotpoint}}
\multiput(442,503)(1.686,20.687){9}{\usebox{\plotpoint}}
\multiput(457,687)(1.996,20.659){8}{\usebox{\plotpoint}}
\multiput(474,863)(2.246,20.634){7}{\usebox{\plotpoint}}
\multiput(490,1010)(3.109,20.521){5}{\usebox{\plotpoint}}
\multiput(505,1109)(7.389,19.396){2}{\usebox{\plotpoint}}
\put(525.51,1146.49){\usebox{\plotpoint}}
\multiput(536,1136)(5.177,-20.099){3}{\usebox{\plotpoint}}
\multiput(553,1070)(3.098,-20.523){6}{\usebox{\plotpoint}}
\multiput(569,964)(2.454,-20.610){6}{\usebox{\plotpoint}}
\multiput(584,838)(2.516,-20.602){6}{\usebox{\plotpoint}}
\multiput(600,707)(2.662,-20.584){6}{\usebox{\plotpoint}}
\multiput(615,591)(3.937,-20.379){4}{\usebox{\plotpoint}}
\multiput(632,503)(6.213,-19.804){3}{\usebox{\plotpoint}}
\put(662.88,443.07){\usebox{\plotpoint}}
\put(672.70,461.19){\usebox{\plotpoint}}
\multiput(679,473)(5.488,20.017){3}{\usebox{\plotpoint}}
\multiput(696,535)(3.607,20.440){4}{\usebox{\plotpoint}}
\multiput(711,620)(3.519,20.455){5}{\usebox{\plotpoint}}
\multiput(727,713)(3.449,20.467){4}{\usebox{\plotpoint}}
\multiput(742,802)(4.386,20.287){4}{\usebox{\plotpoint}}
\multiput(758,876)(6.681,19.651){2}{\usebox{\plotpoint}}
\multiput(775,926)(12.064,16.889){2}{\usebox{\plotpoint}}
\put(804.27,939.87){\usebox{\plotpoint}}
\put(813.60,921.78){\usebox{\plotpoint}}
\multiput(821,905)(5.998,-19.870){3}{\usebox{\plotpoint}}
\multiput(837,852)(5.252,-20.080){3}{\usebox{\plotpoint}}
\multiput(854,787)(4.600,-20.239){4}{\usebox{\plotpoint}}
\multiput(869,721)(5.432,-20.032){2}{\usebox{\plotpoint}}
\multiput(885,662)(6.697,-19.645){3}{\usebox{\plotpoint}}
\put(908.33,604.46){\usebox{\plotpoint}}
\put(921.96,590.60){\usebox{\plotpoint}}
\put(939.66,594.66){\usebox{\plotpoint}}
\multiput(948,603)(9.282,18.564){2}{\usebox{\plotpoint}}
\multiput(964,635)(6.836,19.597){2}{\usebox{\plotpoint}}
\multiput(979,678)(6.689,19.648){2}{\usebox{\plotpoint}}
\multiput(995,725)(7.335,19.416){3}{\usebox{\plotpoint}}
\multiput(1012,770)(7.621,19.306){2}{\usebox{\plotpoint}}
\put(1036.35,822.61){\usebox{\plotpoint}}
\put(1050.00,837.67){\usebox{\plotpoint}}
\put(1068.85,840.45){\usebox{\plotpoint}}
\put(1084.89,828.49){\usebox{\plotpoint}}
\put(1096.75,811.65){\usebox{\plotpoint}}
\multiput(1106,795)(9.055,-18.676){2}{\usebox{\plotpoint}}
\multiput(1122,762)(8.589,-18.895){2}{\usebox{\plotpoint}}
\multiput(1137,729)(10.233,-18.058){2}{\usebox{\plotpoint}}
\put(1164.76,683.53){\usebox{\plotpoint}}
\put(1179.04,668.77){\usebox{\plotpoint}}
\put(1198.02,662.37){\usebox{\plotpoint}}
\multiput(1201,662)(18.808,8.777){0}{\usebox{\plotpoint}}
\multiput(1216,669)(14.676,14.676){2}{\usebox{\plotpoint}}
\put(1244.32,700.85){\usebox{\plotpoint}}
\put(1255.90,718.05){\usebox{\plotpoint}}
\multiput(1264,731)(11.853,17.038){2}{\usebox{\plotpoint}}
\put(1291.76,768.89){\usebox{\plotpoint}}
\put(1307.33,782.43){\usebox{\plotpoint}}
\put(1326.20,790.44){\usebox{\plotpoint}}
\multiput(1328,791)(20.573,-2.743){0}{\usebox{\plotpoint}}
\put(1346.26,787.17){\usebox{\plotpoint}}
\put(1363.48,775.82){\usebox{\plotpoint}}
\put(1378.50,761.50){\usebox{\plotpoint}}
\put(1393.18,746.82){\usebox{\plotpoint}}
\put(1407.86,732.14){\usebox{\plotpoint}}
\put(1422.60,717.55){\usebox{\plotpoint}}
\put(1439.42,705.50){\usebox{\plotpoint}}
\put(1459.23,699.72){\usebox{\plotpoint}}
\put(1479.66,701.42){\usebox{\plotpoint}}
\put(1498.55,709.69){\usebox{\plotpoint}}
\put(1515.82,721.19){\usebox{\plotpoint}}
\put(1532.78,733.14){\usebox{\plotpoint}}
\multiput(1534,734)(16.207,12.966){0}{\usebox{\plotpoint}}
\put(1549.05,746.03){\usebox{\plotpoint}}
\put(1567.22,756.04){\usebox{\plotpoint}}
\put(1586.60,762.83){\usebox{\plotpoint}}
\put(1607.27,763.34){\usebox{\plotpoint}}
\put(1627.51,759.13){\usebox{\plotpoint}}
\multiput(1628,759)(19.015,-8.319){0}{\usebox{\plotpoint}}
\put(1646.38,750.57){\usebox{\plotpoint}}
\put(1664.40,740.30){\usebox{\plotpoint}}
\put(1683.06,731.21){\usebox{\plotpoint}}
\put(1702.10,722.96){\usebox{\plotpoint}}
\put(1722.21,718.15){\usebox{\plotpoint}}
\multiput(1723,718)(20.756,0.000){0}{\usebox{\plotpoint}}
\put(1742.92,718.61){\usebox{\plotpoint}}
\put(1763.33,722.20){\usebox{\plotpoint}}
\put(1782.95,728.78){\usebox{\plotpoint}}
\multiput(1786,730)(19.434,7.288){0}{\usebox{\plotpoint}}
\put(1802,736){\usebox{\plotpoint}}
\end{picture}
\end{center}
 \caption{Time evolution of the S\&P 500 index over a
time window of a few weeks during and after the crash of October
19, 1987. The fit with an exponentially decaying sinusoidal
function suggests that a good model for the short-time response of
the US market is close to {\it single} dissipative harmonic
oscillator. Reproduced from (Sornette et al., 1996).}
\label{figcrash}
\end{figure}
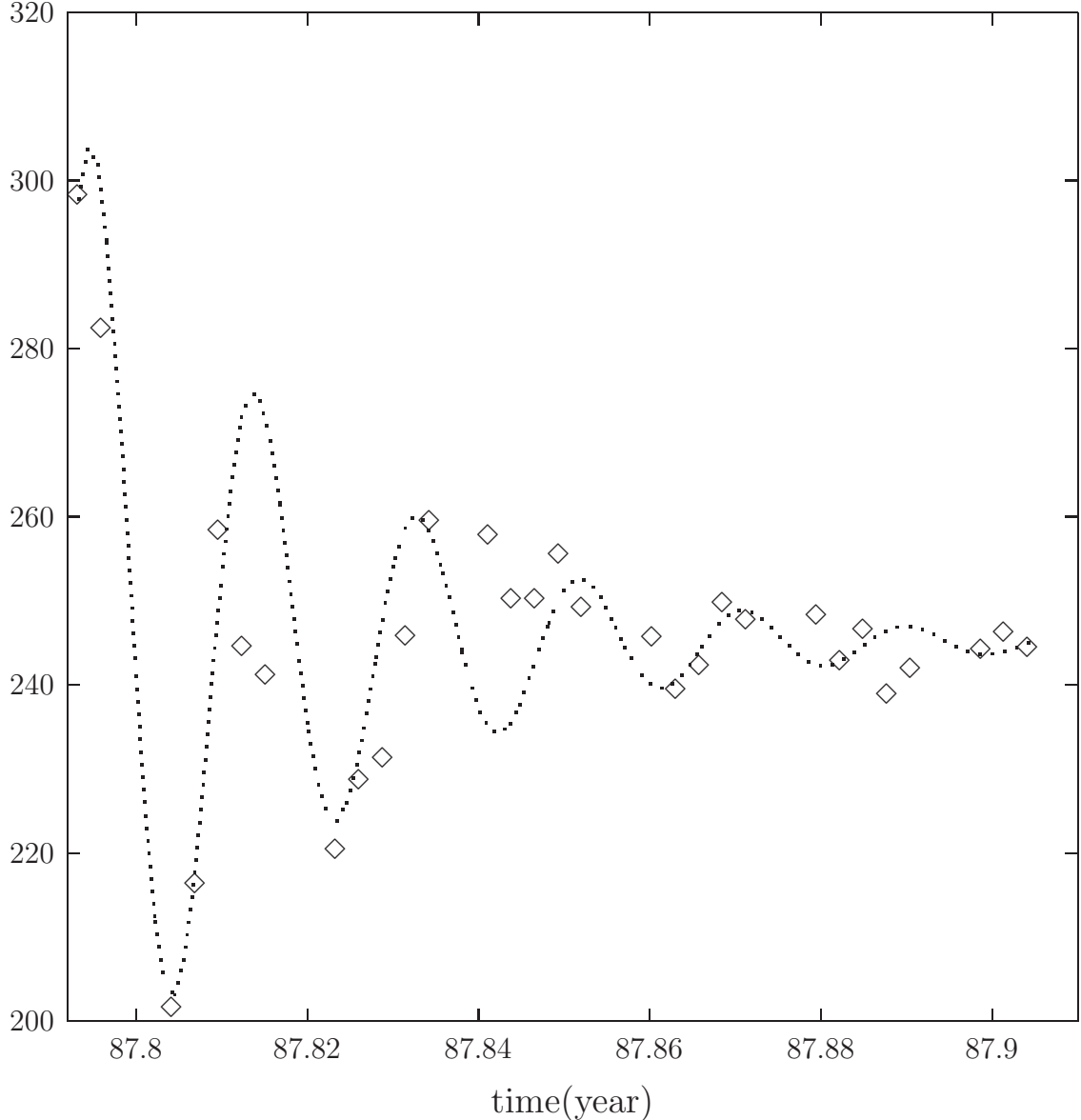

 As a response to the sudden strong `sell' mood in the market, the stock
 price evolution drops
 sharply and, with diminishing selling pressure, approaches in an
 oscillatory manner a new stable value. This is reminiscent of a sudden
 large sell-off of some company stock which unexpectedly publishes some
 bad news but, where due to insufficient conformity or herding pressure
 and an immediate reaction opposing the irrational sell-off, the damage
 is limited to a price drop to a new stable though lower level representing
 the new price discounting the novel piece of news. This
 behavior of a sharp drop followed by damped oscillations converging
 to a well-defined lower plateau is also observed very clearly
 for the crash of October 1987, for instance in the US
(see figure 3 of (Sornette et al., 1996) reproduced here as figure
\ref{figcrash}). Symmetrically, if the initial condition for
$y(0)$ is positive, the average opinion $y(t)$ decreases sharply
to $0$ and there is an increase of the stock price level.

To compare with empirical data, one needs to re-introduce a
stochastic component, for instance in the form of an additive
white noise to the right-hand-side of expression (\ref{11})
(Weidlich, 2000). In this case, the fact that $(0,0)$ is stable
corresponds to the log-price $p(t)$ following a random walk.

\subsubsection{Pattern B: ``the CEO's dream''}

\begin{figure}[htb]
\begin{center}
\includegraphics[width=7cm]{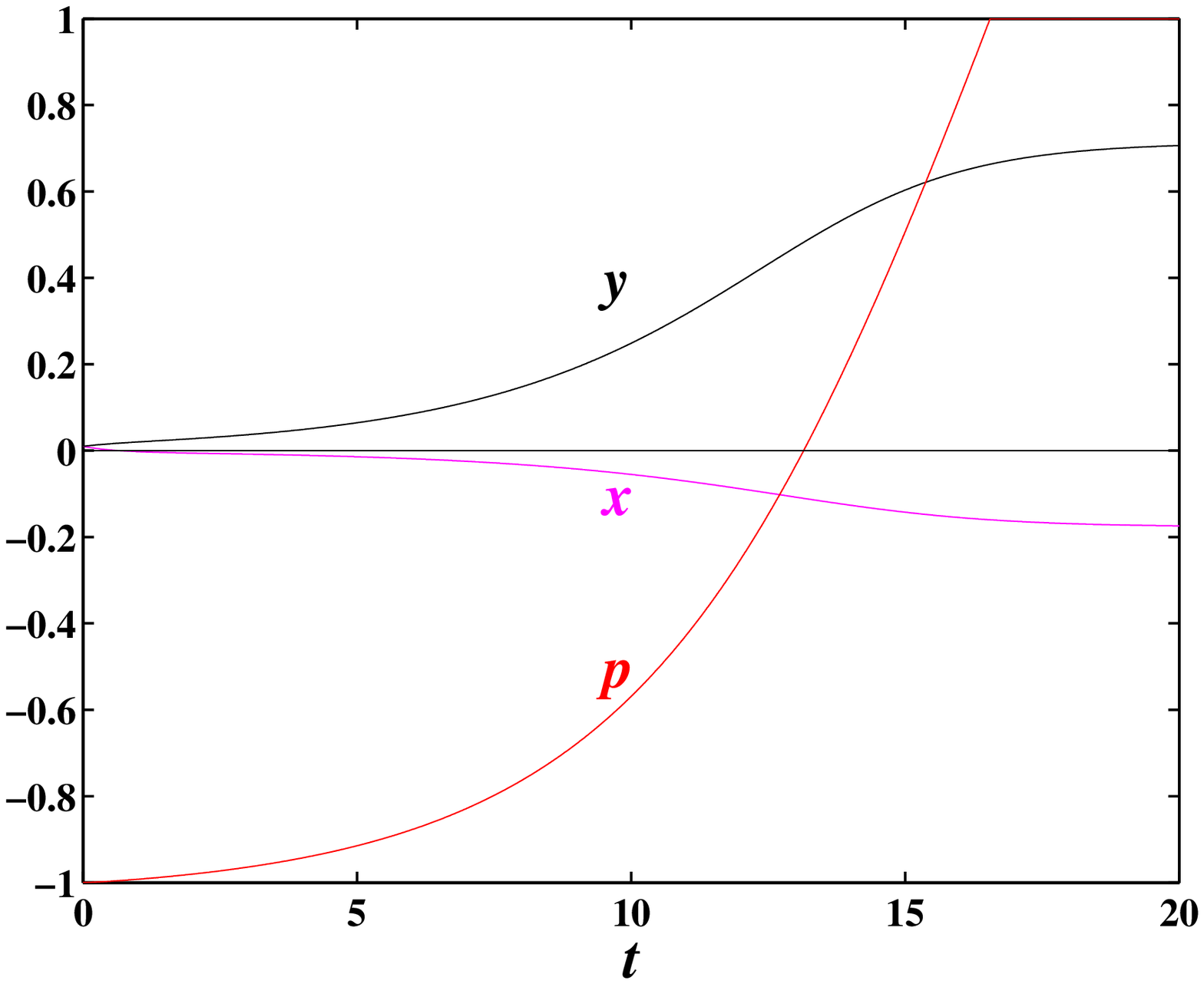}
\end{center}
\caption{Evolution Pattern B: Illustration with $\mu = 2$; $\gamma
= 1$; $\kappa = 1.5$; $\beta = -0.25$; $a = 0.5$. Since $\kappa =
1.5 < 1+\mu = 3$ and $\beta = -0.25 > (1-\kappa)/\gamma =
 -0.5$, condition (\ref{6}) is broken and $(x,y) = (0,0)$ is unstable.
The limit cycle condition (\ref{8}) does
not apply since $b = -0.25 > -(\mu + \kappa -1)^2/(4\mu \gamma) = -0.8$.
Note the upward curvature of the log-price $p(t)$ exemplifying the
super-exponential growth discussed in the text.
 }
\label{Fig2}
\end{figure}

As shown in Figure \ref{Fig2}, this is an interesting situation
because the opposing tendency $\beta$ is too weak to prevent a
runaway situation to occur. In the stock market, this represents
the bullish formation of a bubble as the price trajectory
indicates. In terms of investors' behavior, the value analysts
lose to the trend followers. Eventually, this leads to a new
stable state for $y$ where the buyers form a large majority, while
simultaneously personal preference settles at a negative
stationary value which is not big enough to defuse or destabilize
the bubble. Weidlich (2000) discussed this scenario in the context
of the transition of a liberal to a totalitarian state. He called
it quite appropriately `the dictator's dream,' because while the
pressure to conform is moderate, due to a too small opposition of
dissident opinion, it is enough to transform a liberal situation
into a completely totalitarian state. Likewise, in the stock
market, this situation of a bullish market could be called the
CEO's dream.

A more precise analytical understanding of this regime can be
obtained by neglecting $x$ in expression (\ref{y1eq}). Starting
with a small initial value $y(0)$, the initial growth of the
global opinion follows
\begin{equation}
\frac{d y}{dt} \approx \kappa y + \frac{1}{6} \kappa^3 y^3 - y
\left(1+ \frac{1}{2} \kappa^2 y^2 \right) + {\rm O}\left( y^5
\right) = (\kappa -1) y +  \left(\frac{\kappa}{ 3} -1\right)
\frac{\kappa^2}{2} y^3 + {\rm O}\left( y^5 \right)~, \label{fgell}
\end{equation}
up to third order in $y$.  As long as $y$ is small, it first grows
exponentially (corresponding to the first linear term $(\kappa -1)
y$) for $\kappa >1$. This first regime translates also into an
initial exponential growth of the log-price $p(t)$ and thus to an
exponential of an exponential growth of the price. This is a clear
super-exponential growth characterizing a bubble (Sornette and
Johansen, 2001; Sornette, 2003; Johansen and Sornette, 2004). Such
exponential of an exponential law, proposed to describe phases of
accelerated hyperinflation, has been shown to be essentially
indistinguishable from power law growth leading to an apparent
finite-time singularity of the form (\ref{Eq:PL}); actually, the
discretized version of an accelerated growth leading to
(\ref{Eq:PL}) can been shown to be an exponential of an
exponential growth (Sornette et al., 2003).

Then, as $y$ grows, the nonlinear term starts to dominate and
leads to an even faster transient super-exponential growth if
$\kappa>3$. For instance, if the cubic nonlinearity held
throughout the dynamics with $\kappa >3$, this would give rise
eventually to a solution at later times of the form $y(t) \sim
1/(t_c -t)^{1/2}$ clearly expressing the super-exponential
behavior characteristic of bubbles. In this case, the log-price
would becomes of the form (\ref{Eq:PL}) with $m=1/2$ by
integration of (\ref{11}). However, as $y(t)$ increases and
becomes a finite fraction of $1$, the approximation (\ref{fgell})
becomes inaccurate as $y$ saturates and converges to the stable
fixed point $y^* < 1$, solution of $\sinh (\kappa y) - y ~\cosh
(\kappa y)=0$. A constant asymptotic value of $y$ then translates
into a linear growth of the log-price $p(t)$ and thus to an
exponential growth of the price. This suggests that the
super-exponential characteristic of a bubble is, according to
Weidlich's theory, only a transient regime.

There are additional factors to consider. In addition to the
saturation mechanism captured by the second term $y ~\cosh (\kappa
y + \gamma x)$ in the r.h.s. of (\ref{y1eq}), the
super-exponential growth of the stock price may be halted in many
cases before the new stationary state of $y$ is reached where a
large majority of investors are buyers. As shown below, at some
time during the formation of the stock price bubble, although
initially supportive, some investors will personally become
increasingly worried despite the herding pressures, and the
contrarian inclination $\beta$ is likely to become more strongly
negative. Likewise, the totalitarians will become more bullish and
$\kappa$ may also increase. This emerging battle between the
degree of personal opposition $\beta$ and the degree of public
conformity pressures $\kappa$, between order and disorder, will
eventually induce a critical phase of destabilization of the
bubble. Note that this reasoning amounts to add positive feedbacks
of the opinion and price dynamics on the control parameters of the
contrarian inclination $\beta$ and public conformity pressures
$\kappa$.

Finally, note that pattern B reproduces the well-known lock-in
through fluctuation situation in cases where, rather than two
opposing opinions, two competing new products or technologies
through some random fluctuations in market share declare one
winner, who takes it all, and one loser (Arthur, 1994; Broekstra,
2002). In other words, if in Figure \ref{Fig2} the initial
condition would have been slightly negative rather than positive,
the symmetrical situation of a stock market crash would have been
developed resulting in a new stationary state of a high majority
of sellers. This sensitivity to initial conditions is also
captured by Ising-like models of opinion formation (Sornette,
2003, Fig.4.10) associated with the presence of a critical point
around which the collapse into one of two states bifurcates
essentially randomly.

\subsubsection{Pattern C (``critical zone''): the destabilization
regime ending in a collapse}

\begin{figure}[htb]
\begin{center}
\includegraphics[width=7cm]{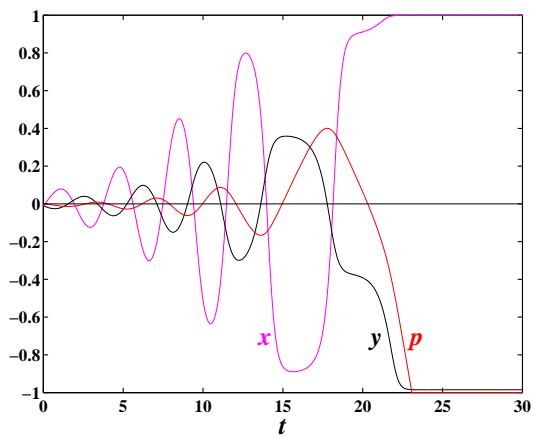}
\end{center}
\caption{Evolution Pattern C: Illustration with $\mu = 2$, $\gamma
= 1$, $\kappa = 3.5$, $\beta = -4$, and $a = 0.5$. Since $\kappa =
3.5 > 1+\mu = 3$ and $\beta = -4 < (1-\kappa)/\gamma =
 -2.5$, condition (\ref{6}) is broken and $(x,y) = (0,0)$ is unstable.
The limit cycle condition (\ref{8}) applies
since $b = -4 < -(\mu + \kappa -1)^2/(4\mu \gamma) = -2.5$.
 }
\label{Fig3}
\end{figure}

Rather than an abrupt transition point, there exists a transition
range between relatively small negative values of $\beta$ below
the limit cycle condition ($-2.5 < \beta < 0$), where $y$ swiftly
approaches one of two opposing new stationary states, and
relatively high negative values of $\beta <- 4.1$, where a true
limit cycle and a stable attractor appears (see pattern D below).
This transition range reflects the dependence of the exact pattern
of evolution on the initial conditions. Within this transition
range $-4.1 < \beta < -2.5$, starting from low values of $|\beta|$
and small initial conditions for $y$, with increasing $|\beta|$,
one observes an increasing number of oscillations of increasing
amplitude which all end in a breakdown characterized, depending on
the initial conditions, by one of two opposing stationary states
of large positive or negative $y$.

Figure \ref{Fig3} shows a typical Pattern C in which, starting
from small initial conditions, both $x$ and $y$ show increasingly
larger and opposing oscillations until a sudden breakdown occurs.
This typical pattern where the stationary state $(0,0)$ has become
unstable and whose unstable oscillatory growth ends in a collapse
deserves the name ``destabilization regime.'' The stock price
shows a similar pattern of increasing oscillations ending in a
crash. This destabilization regime which acts a precursor to an
inevitable collapse only occurs under the condition of strong
herding pressures, $\kappa = 3.5 >  1 + \mu$. This is in
qualitative correspondence with previous observations of the
ubiquitous existence of oscillations of increasing amplitudes
preceding crashes, which have also been interpreted as the
signature of strong herding pressure (Sornette et al., 1996;
Sornette and Johansen, 2001; Sornette, 2003). The tendency to
strongly conform with the prevailing majority is fueled by
exogenous factors like the overall market mood, as happened during
the bull market of the late 1990s, but also endogenous factors
like announcements of acquisitions and management promises of
significantly increased earnings. As we will see below, Ahold is a
case in point.

\begin{figure}[htb]
\begin{center}
\includegraphics[width=7cm]{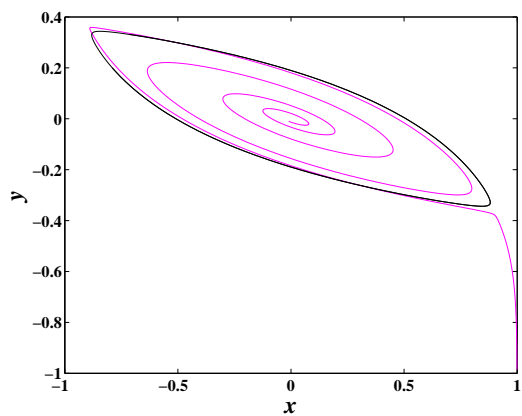}
\end{center}
\caption{Phase portrait in the $x(t); y(t)$ phase space,
corresponding to Figure \ref{Fig3}. The stable limit cycle for
$\beta =-4.2$ is shown as the elliptic-like attractor. The
spiraling trajectory is for $\beta = -4$, which falls in the
destabilization regime. It starts near the origin and escapes the
basin of attraction at the lower right end, corresponding to the
collapse discussed in the text.
 }
\label{Fig4}
\end{figure}

A better understanding of Pattern C can be obtained from its phase
portrait in the $\{x(t), y(t)\}$ phase space. Figure \ref{Fig4}
shows the stable elliptic-like attractor for a slightly higher
negative value $\beta =-4.2$, other parameters being equal. When
started from conditions close to the unstable stationary state
$(0,0)$, all trajectories stay within the domain of attraction and
converge towards it. This leads to a stationary limit cycle (see
Pattern D below). Note that the attractor is somewhat deformed at
its far ends, tipping slightly upward on the far left side, and
downward on the far right side. Increasing $\beta$ slightly to
$-4$, as shown above, enters the transition range characterized by
trajectories starting close to $(0,0)$ as shown in Figure
\ref{Fig4}: after a number of increasing oscillations, the
trajectory escapes at the far end from the attractor to collapse
in the lower right bottom of the $(x,y)$ phase space. Under
different initial conditions, the trajectory may also escape from
the upper far end on the left side to the upper right corner of
the phase space.

\subsubsection{Pattern D: `Roller coaster'}

\begin{figure}[htb]
\begin{center}
\includegraphics[width=7cm]{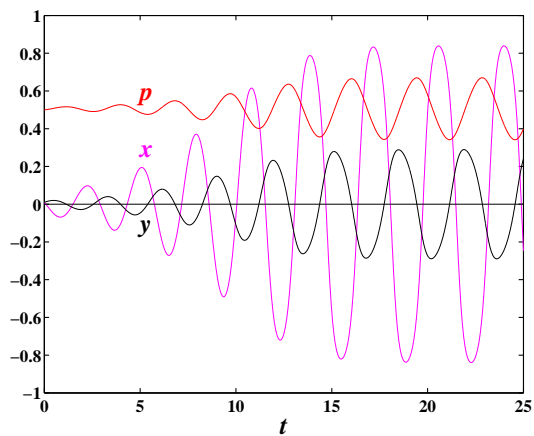}
\end{center}
\caption{Evolution Pattern D: the Roller Coaster. Starting from
small initial conditions near $(0,0)$, the trajectories develop
into stable sinusoidal patterns. The simulation shown corresponds
to $\mu = 2$, $\gamma = 1$, $\kappa = 3.5$, $\beta = -5$, $a = 1$.
 }
\label{Fig5}
\end{figure}

This regimes corresponds to large negative $\kappa$ and $\beta$, such
that $(0,0)$ is unstable but a stationary limit cycle exists, as shown in
Figure \ref{Fig5}. The phase portrait of this limit cycle corresponding to
Figure \ref{Fig5} is similar to that shown in Figure \ref{Fig4} for $\beta=-4.2$
but with more rounded edges, which explains the
smoothed regular sinusoidal movement shown in Figure \ref{Fig5}.

It is somewhat unlikely that such a pattern of big upswings and
downswings from a majority of buyers to a majority of sellers and
vice versa and corresponding `roller coaster' swings of the stock
price will be sustainable for a long time. It indicates extreme
uncertainty among investors about the company's predicament. As
shown, it can only occur when the inclination $\beta$ to
disapprove of the prevailing majority remains extremely strong,
causing big oscillations in personal preference. However, the
slightest drop of $|\beta|$ will cause the system to fall back
into the destabilization regime ending in a collapse (Pattern C).

\subsubsection{Summary of the classification of patterns}

Let us now summarize the obtained classification in four
fundamental patterns of evolution. As long as the conformity
parameter $\kappa$ remains sufficiently small and less than
$1+\mu$, indicating little propensity among investors to herd, and
as long as the inclination parameter $\beta$ remains also below
its threshold $(1- \kappa)/\gamma$, counteracting as it were any
inclination to herd, the stability condition (\ref{6}) for $(0,0)$
applies and nothing much interesting happens. This is Pattern A.
However, if the stability condition (\ref{6}) for $(0,0)$ breaks
down, either the herding pressure exceeds the critical value
$1+\mu$, or if it does not but the counteracting disapproval
parameter $\beta$ is insufficiently strong, the investors' system
locks into a runaway situation. Whether this is a bubble or a
crash sensitively depends on the initial conditions. This is
evolution Pattern B.

If the stability condition (\ref{6}) breaks down, and for a given
herding pressure $\kappa$, when the inclination parameter $\beta$
becomes sufficiently negative so that condition (\ref{8}) applies,
a broad transition range of $\beta$ values appears between
unstable growing oscillations ending in a run-away (Pattern C) for
smaller values of $|\beta|$ and a stable limit cycle for
sufficiently large $|\beta|$ (Pattern D).  Pattern C is the more
interesting pattern particularly when the bullish rise of the
stock price is temporarily halted in a moment of hesitation, and
according to Eq.(\ref{11}), $y(t)$ becomes equal to zero. Within
the transition range, it appears as if a climate of escalation
between trend followers and value investors induces a
destabilization regime which acts as a precursor to an inevitable
collapse (Sornette et al., 1996; Sornette and Johansen, 2001;
Sornette, 2003).

The specifics of the basic evolution patterns are also governed by
the speed and influence parameters $\mu$ and $\gamma$. To simplify
the discussion, we have kept them constant throughout.

As already indicated, in actual practice, the conformity and
personal inclination parameters $\kappa$ and $\beta$ will not
remain constant but will change under the influence of or
co-evolve with changes of $x(t)$ and $y(t)$. This would require
the introduction of evolution equations for these parameters, thus
complicating the model considerably not in the least because it
will also introduce aspects of chaos dynamics. A simpler approach,
which has the advantage of bringing out the essence of what is
involved, is to change parameters discontinuously. Weidlich and
Haag (1983) has successfully applied this method in reproducing
the strategic investment evolution in Germany (see also Weidlich,
2000). We have followed this approach to reproduce the qualitative
and main features of the Ahold stock price evolution. In this way,
we propose an interpretation and understanding of what happened.

\section{Interpretation of the stock price evolution of Ahold}

\begin{figure}[htb]
\begin{center}
\includegraphics[width=7cm]{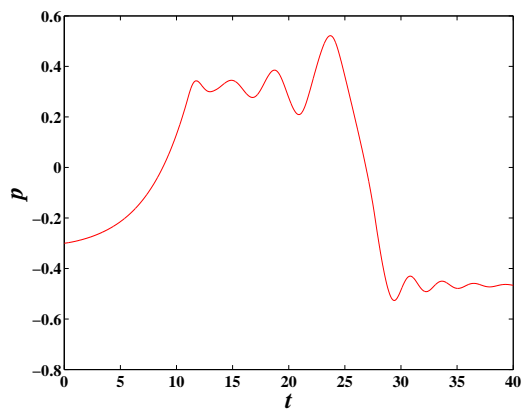}
\end{center}
\caption{This figure shows a simulation of the log-price using
Weidlich's equations (\ref{y1eq}) and (\ref{x1eq}) performed to
reproduce the basic qualitative features of the Ahold stock price
evolution shown in Figure \ref{Fig:33}.
To obtain the corresponding log-price
trajectory, four stages can be distinguished in which each time
just one parameter was changed discontinuously. The parameters
$\mu = 2$, $\gamma = 1$, and $a= 0.5$ are kept constant throughout
the simulation. Only $\kappa$ and $\beta$ are changed.
 }
\label{Fig6}
\end{figure}

Figure \ref{Fig6} shows the result of a simulation using
Weidlich's equations (\ref{y1eq}) and (\ref{x1eq}) performed to
reproduce the basic qualitative features of the Ahold stock price
evolution. For this, four phases are distinguished, each
characterized with a different pair $(\kappa, \beta)$.

\subsection{Phase 1 (time period: $0 \leq t \leq 12$, i.e. 1993-mid 1997):
$\kappa = 1.5, \beta = -0.25$ (Pattern B)}

This regime corresponds to Pattern B, in which a bullish
super-exponential price rise occurs because of relatively moderate
herding pressures which are not counteracted by a sufficiently
strong personal disapproval. This is the CEO's dream. From roughly
1995/1996 onwards, the stock market as a whole (the AEX) had
become increasingly bullish. So Ahold was riding the bull when it
accelerated its growth strategy during that period. In the third
quarter of 1996, due the consolidation of the large American
supermarket chain Stop \& Shop, the quarterly net earnings
increase, which during the preceding period in the 1990s usually
remained well below 20\%, discontinuously rose to 54\%, initiating
a new period of (promises of) considerably higher increases in
both sales and earnings.

\subsection{Phase 2 (time period: $12 < t \leq 14$, i.e., mid 1997-end 1997):
$\kappa = 1.5,  \beta = -4$ (Pattern A)}

At some point in time for some reason, investors hesitate and the
stock price reaches a first peak. Perhaps fear of heights and/or a
decrease in the stock market as a whole (the AEX showed a peak
followed by a temporary decline in mid 1997) have increased
investors' wariness and fueled the contrarians. Just by decreasing
$\beta$ from a weak ($\beta= -0.25$) to a more strongly
disapproving propensity ($\beta= -4$), the herding pressure is
temporarily counteracted such that the stability condition for
$(0,0)$ applies and Pattern A occurs (actually the inverse of the
example given in Fig.\ref{Fig1}). After a peak in the stock price,
when $y(t) =0$ (see Eq.(\ref{11}), the bullishness quickly returns
with a vengeance leading to phase 3.

\subsection{Phase  3 (time period: $t > 14$ until $p(t)$ drops below $p(0)$, i.e.,
1998- end 2002): $\kappa = 3.5, \beta = -4$ (Pattern C)}

Keeping the degree of dissidence high, $\beta = -4$, but now
increasing the herding pressures from $\kappa = 1.5$ to $3.5$,
indicating both the general bullish market tendencies and
investors' expectations about the company, Pattern C comes into
play. By the end of 1997, Ahold had completed its first peak and a
new upward trend occurred. This point in time could be regarded as
the start of the destabilization regime which became increasingly
clear over time when further oscillations occurred. This was the
precursor for the collapse of the stock price in 2002 followed by
a final blow due to the publication of fraud in February 2003.

In general, it is not unlikely that top management fuels the
dynamics of Pattern C, by raising expectations of increased
earnings in the investors' community. For example, before 1996,
the year the growth rate accelerated, the announcement that
`further increases' in earnings (per share) were expected was the
standard statement in press releases of Ahold. After the report of
the unprecedented 54\% earnings increase in the third quarter of
1996, announcements were upgraded in terms of `significantly' or
`considerably higher earnings' that could be expected. Finally,
the company became daringly precise and bullish by announcing as
of March 1999 that its outlook for the whole year foresaw an
expected earnings per share growth of at least 15\%.

However, 1999 was also the year that Ahold's stock price declined
44\% from an all time high of 38.5 euro in April to a low of 21.2
euro in February 2000, while in the same period the Dutch stock
market index AEX showed an overall upward trend and appreciated by
more than 20\%. The start of this downward trend could have made
Ahold's top management somewhat nervous, and in an unprecedented
statement, the company included an upbeat quote of its CEO in its
10 June 1999 press release, covering the Q1 1999 results and
outlook for 1999, saying that the prospects were excellent and
that the company was well on course. A statement like this could,
of course, fuel the conformity pressures towards bullishness.

On the other hand, a growing skepticism may have developed among
investors who were concerned by the overvaluation of the stock and
disapproved of a change in management strategy. This happened in
1999 to Ahold, after the failed takeover of yet another
supermarket chain (Pathmark in the USA),  when it announced the
acquisition of US Foodservice at the end of 1999. Some analysts
(including Standard \& Poor's) disapproved of this change of
course to the unfamiliar non-core activities of the wholesale food
service industry. Also, early 2001, the dollar started its
decline, which may have increased investors' degree of disapproval
throughout the last part of the destabilization phase. This
situation of increasing opposing tendencies is reflected in the
relatively high values of the conformity parameter $\kappa$ and
the negative inclination parameter $\beta$, which are responsible
for the destabilization as a precursor to the unavoidable collapse
of the stock price as shown above in Pattern C and Figure
\ref{Fig6}.

The year 1999 appears to have been a critical year. With the
destabilization well on its way (according to our classification),
Ahold's stock price continued its overall decline in 1999 until
the higher increase of net earnings of 37\% in 1999 (compared with
29\% in 1998) was reported in March 2000. This together with the
future consolidation of US Foodservice and other acquisitions in
2000 (other food service companies and PYA/Monarch in the USA
among others) in 2000 were expected to further enhance growth. By
then, the AEX was also on a final growth spurt which ended early
September 2000 when the market started its long decline. The Ahold
stock price then started its last mini-bubble, undoubtedly fueled
by these expectations of higher earnings. As became evident in
2003, however, these turned out to be misplaced, and the company
performance results had to be corrected downwards for 2000 and
2001.

Qualitatively, the overall oscillating pattern of stock price
evolution after the first peak is strikingly similar to that of
Pattern C. Although Ahold price shows a somewhat increasing trend
in this regime, it should be noted that, for reasons of clarity,
longer-term trend lines have been left out of the above simple
model of patterns of evolution (see also Weidlich's example of
Schumpeter's clock, 1983).

Pattern C shows that the increasing oscillations are a part of the
destabilization of the bubble, which, if not defused in time, is
the precursor for an impending collapse as happened with Ahold.
Pattern C is really the expression of the age-old adage: What goes
up, must come down.

\subsection{Phase  4 (Beyond the time where $p(t)$ drops below $p(0)$: 2003- mid 2003):
$\kappa = 2.5, \beta = -4$ (Pattern A)}

Some time during the collapse of the stock price, here somewhat
arbitrarily located at the point in time when $p(t)$ drops below
$p(0)$, the herding pressures will deflate, and we are back to
Pattern A. In the process, the inclination parameter $\beta$ will
also return to smaller values, but to keep to the essentials, for
a return to the stability of $(0,0)$, it is already sufficient for
the conformity parameter to drop below $1+\mu$.

With phase 4, we have completed the description of a complete
bubble using a very simple synergetics model and Ahold as an
example to demonstrate the power of the method. The discovery of
the destabilization regime after a bullish rise of the market
enhances the possibility of developing early warning signals by
focusing on pattern recognition rather than on short-term gains.
Would this knowledge have been available, in Ahold's case, having
spotted its growth acceleration in 1996 to super-exponential
proportions, as early as the end of 1997 when the first
oscillation was completed and the next one started, investors
could have become somewhat concerned about the consequences of
their own bullishness by realizing that they were really riding
the waves of a self-created destabilization regime which, if
nothing changed, is the precursor of an inevitable collapse,
sooner or later. But certainly by the end of 1999, the
destabilization pattern C was quite evident and an eventual
collapse could only have been prevented by defusing overheated
expectations. This evidently did not happen, on the contrary, and
the destabilization regime ran its course ending in a dramatic
collapse of the stock price.

The paradoxical conclusion is that, particularly at times of
overheated post-bubble stock markets where strong herding
pressures are counteracted by equally strong contrarian
propensities, companies are advised not to throw oil on the fire
by overly raising or even manipulating expectations of higher
earnings. Although it may go against the grain for ambitious CEOs,
given the humans' latent animal instinct to herd, cooling the
stampeding crowd of investors would be a far better policy in an
attempt to defuse a dangerous destabilization regime which acts as
the precursor of a future collapse.

Destabilization is not an abstract phenomenon that is without
consequences for the internal functioning of a company. Although
this aspect requires further research, it is highly likely that in
the case of Ahold starting in 1999 a certain loss of control
became internally visible. For example, newspapers have reported
that the bookkeeping problems at US Foodservice were already known
in 2000. Dogmatic preoccupation with the growth strategy by top
management, lack of adequate supervision by and short-term
orientations of directors and accountants alike prevented the
internal consequences of the destabilization to surface in time.

\subsection{Another corporate example: Aegon}

\begin{figure}[htb]
\begin{center}
\includegraphics[width=7cm]{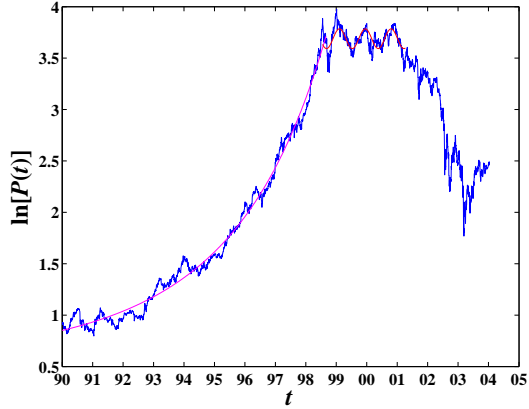}
\end{center}
\caption{Aegon stock price as a function of time. The vertical
axis uses a logarithmic scale while the horizontal axis uses a
linear scale. The power-law fit to the data from $\rm{1990/01/02}$
to $t_{\rm{max}}=\rm{1998/07/21}$ gives $t_c={\rm{2015/06/15}}$,
$m=-5.40$, $A=0.51$, and $B=9.48\times10^{20}$ with a r.m.s. of
the fit residuals equal to $\chi=0.078$. The sinusoidal fit to the
data from $\rm{1998/07/21}$ to $\rm{2001/03/20}$ gives
$\omega=0.0204$, $\phi=3.73$, $A=3.69$, $B= -0.0959$ with a r.m.s.
of the fit residuals equal to $\chi=0.0755$.} \label{FigAegon}
\end{figure}

The Dutch insurance group Aegon is one of the largest in the world
operating in three major markets of North-America, The Netherlands
and the United Kingdom. Its net income for the year 2003 was 1.8
billion Euros of which, like Ahold, the largest share is in the
USA. Figure \ref{FigAegon} shows another remarkable illustration
of a similar sequence of phases as discussed for Ahold. The first
regime is characterized by an upward curvature of the log-price of
Aegon as a function of time, which is again a signature of
super-exponential growth, that culminated mid-1998. Then, a
plateau decorated with oscillations from mid-1998 to the beginning
of 2001 can be represented rather faithfully by the
``roller-coaster'' pattern D (fitted with a regular sinusoidal
oscillation $\ln[P(t)] = A+B\cos(\omega t + \phi)$). This second
phase contrasts with the enthusiasm of investors of the first
phase, as being a time of large uncertainty among investors about
the company's future. Recall that this regime only occurs when the
personal inclination $\beta$ to disapprove of the prevailing
majority remains extremely strong. Then, the price seems to
transition from pattern D to the critical zone C ending in a
crash. The subsequent behavior is reminiscent of pattern A in
which oscillations are progressively dampened towards an
equilibrium point.

\begin{figure}[htb]
\begin{center}
\includegraphics[width=7cm]{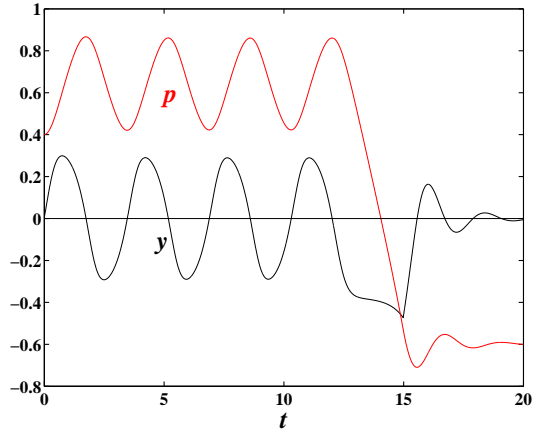}
\end{center}
\caption{Simulation of the stock price pattern of Aegon after the
bubble with $\mu=2$, $\gamma=1$, and $a=1$.} \label{Fig7}
\end{figure}

Figure \ref{Fig7} shows a schematic synthesis of the regimes after
the bubble phase (Phase 1 and 2) which is basically similar to the
Ahold simulation.

{\bf{Phase 3 (time period  $0 \le t \le 12$):  $\kappa = 3.5$ and
$\beta = -5$.}} This is pattern D, where after the bubble there
are strong herding pressures and such strong dissident
inclinations that $\beta$ is beyond the transition range, and a
stable limit cycle results. However, as suggested when we
discussed Pattern D above, if the disapproval starts to relax
somewhat so that $\beta$ falls back into the transition range, the
stable wave quickly destabilizes and collapses. Although due to
very strong herding and dissidence pressures a stable limit cycle
is produced, the danger lures that, with the slightest relaxing of
the pressures, the system is likely to fall back into the
destabilization regime causing it to collapse.

{\bf{Phase 4 (time period  $12 < t \le 15$): $\kappa = 3.5$,
$\beta = -4.1$ (Pattern C).}} With $\beta$ now in the transition
range, the limit cycle destabilizes and, as can be seen from Fig.
\ref{Fig7}, the $y$ trajectory starts to collapse causing a stock
price crash. This would continue indefinitely, so like in the
Ahold case, when investors become dizzy from falling so fast, the
herding pressures will evaporate, and the balance between buyers
and sellers will return.

{\bf{Phase 5 (time period  $t > 15$):  $\kappa = 1.5$, $\beta =
-4.1$ (Pattern A).}} This is the return to the $(0,0)$ stability
zone, as exemplified in Figure \ref{figcrash}.

\section{Concluding remarks}

We have documented the concept of super-exponential growth
characterizing corporate sales, earnings and stock market prices
of companies such as Ahold and Aegon which developed a furious
race for growth via aggressive acquisitions as well as creative
accounting. These behaviors allowed these companies to create
extraordinary positive or bullish sentiments among investors, and
to manipulate these sentiments for the sake of growth feeding
positively on itself. It is well-known that high stock prices help
a company grow and reciprocally the company's growth fuels the
stock prices: a high stock market price makes it easier to raise
money, to acquire other companies, to attract high level
collaborators and employees and so on. As pointed out by Krugman
(2002), a high stock price facilitates aggressive plans for growth
as well as accounting tricks, making the growth similar to a Ponzi
scheme. The innovation of our paper has been to propose a simple
quantification of these unsustainable growth phases often reported
at a qualitative or descriptive level.

We have adapted Weidlich's theory of opinion formation to describe
the formation of buy or sell decisions among investors, based on a
competition between the mechanisms of herding and of personal
opinion opposing the herd. A study of this model has provided a
classification of four different regimes/patterns for the price,
which have then be used to explain the price evolution of Ahold as
well as Aegon, from the bubble phase to the final crash. This
study complements previous studies of bubbles holding that a crash
is the most probable close to the end of a bubble by introducing
the concept of a ``critical zone'' corresponding to a transition
from the super-exponential bubble to the crash. The critical zone
is characterized by (i) herding and (ii) strong shift of
sentiments (or in other words, instabilities in the shift of
sentiments). The critical zone is characterized by a strong
sensitivity on these herding and inclination parameters and
describes the maturation of an instability, that is, the coming
crash. These ingredients have allowed us to explain the
trajectories of Ahold and Aegon by suitable choices of the herding
and inclination parameters, according to scenarios and a
succession of phases which are essentially identical. While we
have focused on just these two cases in order to get sufficient
depth, we conjecture that more studies on other cases will allow
to strengthen and refine our model.

Concerning the potential for prediction, the important point of
our analysis is to focus on medium-term evolutionary patterns
which leads to the possibility of recognizing years in advance
that a critical zone is unfolding following a super-exponential
bubble phase, which foreshadows a coming crash. We have been able
to show from the power law fits of sales and earnings of Ahold
that about a year in advance a $t_c$, which indicates a change of
regime, was predicted, {\it{i.e.}} a crash in 2002. About five
years earlier, in 1997, while sales and earnings were still in the
super-exponential growth regime, the stock price evolutionary
pattern entered a critical zone characterized by a destabilization
regime that over the years became increasingly apparent and was
bound to end in a crash as the remarkable Pattern C of the
critical zone has shown. Our findings that the {\it{combination}}
of these patterns, continuing super-exponential growth of sales
and earnings of a company that resembles an unguided missile, and
a stock market which exhibits the collective intelligence and
transparency to indicate, at least as early as 1998,  that the
missile is out of control and moving through a dangerous
destabilization zone entails an important lesson for top managers
and investors alike.

\textbf{Acknowledgments}

We acknowledge stimulating exchanges with Emeritus Professor
Wolfgang Weidlich of the Institute for Theoretical Physics of the
University of Stuttgart, Germany and help of Carel ter Ellen of
Rabobank International in retrieving some of the publicly
accessible raw data. This work was supported by the James S.
McDonnell Foundation 21st century scientist award/studying complex
system.



\begin{thebibliography}{999}


\bibitem{Abreu} D. Abreu, Bubbles and Crashes,
Princeton Working Paper (2001) (http://ssrn.com/abstract=296701)

\bibitem{Arthur} W.B. Arthur, Increasing returns and path dependence in the economy
(The University of Michigan Press, Ann Arbor, 1994).

\bibitem{Bohl} M.T. Bohl,
Periodically Collapsing Bubbles in the US Stock Market?
International Review of Economics and Finance 12 (2003) 385-397.

\bibitem{Broekstra} G. Broekstra, A Synergetics approach to
disruptive innovation, Kybernetes: The International Journal of Systems
\& Cybernetics 31 (9/10) (2002) 1249-1258.

\bibitem{Brooks} C. Brooks and A. Katsaris,
Rational Speculative Bubbles: An Empirical Investigation of the
London Stock Exchange, Bulletin of Economic Research 55 (2003) 319-346.

\bibitem{Brun} M.K. Brunnermeier and S. Nagel,
Hedge Funds and the Technology Bubble, EFA 2003 Annual Conference
Paper No. 446\\ (http://ssrn.com/abstract=423940)

\bibitem{Caballero} R.J. Caballero and M. Hammour, Speculative Growth,
MIT Department of Economics Working Paper No. 02-45 (2002)\\
(http://ssrn.com/abstract=355901)

\bibitem{Cardy} J.L. Cardy, ed., Finite-size Scaling (North-Holland, Amsterdam, 1988).

\bibitem{Chari} V.V. Chari and P.J. Kehoe,
Financial Crises as Herds: Overturning the Critiques, NBER Working
Paper No. W9658 (2003)\\ (http://ssrn.com/abstract=398561)



\bibitem{Engtan} T. Engsted and C. Tanggaard,
Speculative Bubbles in Stock Prices? Tests Based on the
 Price-Dividend Ratio, SSRN Electronic Paper Collection  (2004)
(http://ssrn.com/abstract=486006).

\bibitem{EHM} M. Erickson, M.
Hanlon and E.L. Maydew, Is There a Link Between Executive
Compensation and Accounting Fraud? (February 24, 2004).
http://ssrn.com/abstract=509505.

\bibitem{Fair} R.C. Fair, Testing for a New Economy in the 1990s,
Cowles Foundation Discussion Paper No. 1388; Yale ICF Working
Paper No. 02-48 (2002) (http://ssrn.com/abstract=363560).

\bibitem{Farmer} J.D. Farmer, Market Force, Ecology and Evolution,
Preprint available at adap-org/9812005 (1998).

\bibitem{Griffin} J. Griffin, J.H. Harris and S. Topaloglu,
Investor Behavior over the Rise and Fall of Nasdaq, Yale ICF
Working Paper No. 03-27 (2003) (http://ssrn.com/abstract=459803).

\bibitem{Haken} H. Haken, Synergetics, an Introduction, 3rd ed. (Springer, Berlin, 1983).

\bibitem{Haken2} H. Haken,
Information and self-organization: A macroscopic approach to complex systems
(Springer, Berlin, 2nd edition, 2000).

\bibitem{Helbing} D. Helbing, A mathematical model for attitude formation by pair
interactions, Behavioral Science 37, 190-214 (1992).

\bibitem{Idesor} K. Ide and D. Sornette,
Oscillatory Finite-Time Singularities in Finance, Population and Rupture,
Physica A  307 (2002) 63-106.

\bibitem{Jiao} T. Jiao, Stock Market Bubble and Earnings
Management: Model and Evidences From China's Stock Market, working
paper from Shanghai Jiao Tong University (2003)
(http://ssrn.com/abstract=474922).

\bibitem{JLS00} A. Johansen, O. Ledoit and D. Sornette,
Crashes as critical points,
International Journal of Theoretical and Applied Finance
3(2) (2000) 219-255.

\bibitem{JS99} A. Johansen and D. Sornette,
Financial ``anti-bubbles'': log-periodicity in Gold and Nikkei collapses,
Int. J. Mod. Phys. C 10 (1999) 563-575.

\bibitem{JS00} A. Johansen and D. Sornette,
The Nasdaq crash of April 2000: Yet another example of log-periodicity
in a speculative bubble ending in a crash,
European Physical Journal B 17 (2000) 319-328.

\bibitem{JS01PA} A. Johansen and D. Sornette, Finite-time
singularity in the dynamics of the world population and economic
indices, Physica A 294 (2001) 465-502.

\bibitem{JS04endo} A. Johansen and D. Sornette,
Endogenous versus Exogenous Crashes in Financial Markets, in press
in ``Contemporary Issues in International Finance'' (Nova Science
Publishers, 2004) (http://arXiv.org/abs/cond-mat/0210509).

\bibitem{JSL99} A. Johansen, D. Sornette and O. Ledoit,
Predicting Financial Crashes using discrete scale invariance,
Journal of Risk 1(4) (1999) 5-32.

\bibitem{Kaizoji} T. Kaizoji,
Inflation and deflation in stock markets, working paper at
http://arXiv.org/abs/cond-mat/0401140 (2004)

\bibitem{Kindleberger} Kindleberger, C.P. (2000)
Manias, panics, and crashes: a history of financial crises (4th ed.  New York: Wiley).

\bibitem{Krugman} P. Krugman, Two, three, many?
The New York Times, February 1, 2002, reproduced in ``The great unravelling''
(Allen Lan, Penguin Books, London, England, 2003), pp. 104-106.

\bibitem{Lamdin} D.J. Lamdin,
Rational Bubbles in Individual Stock Prices and Implications for
Investment Management (2002) (http://ssrn.com/abstract=351861).

\bibitem{Levysol} M. Levy, H. Levy and S. Solomon, The Microscopic Simulation of
Financial Markets: From Investor Behavior to Market Phenomena (Academic Press,
San Diego, 2000).

\bibitem{LuxMar} T. Lux and M. Marchesi,
Scaling and criticality in a stochastic multi-agent
model of a financial market, Nature 397 (1999) 498-500.

\bibitem{Miller} R.M. Miller,
Can Markets Learn to Avoid Bubbles?
Journal of Psychology \& Financial Markets 3 (2002)

\bibitem{Rao} C. Rao, Linear statistical Inference and Its
Applications, Wiley, New York, 1965.

\bibitem{Richar} M.P. Richardson and E. Ofek,
DotComMania: The Rise and Fall of Internet Stock Prices, NYU Stern
Department of Finance, Working Paper No. IN-01-037 (2001)
(http://ssrn.com/abstract=293964)

\bibitem{Xiong} J.A. Scheinkman and W. Xiong,
Overconfidence and Speculative Bubbles,
Journal of Political Economy 111, December 2003

\bibitem{Shefrin} Shefrin, H. (2000)
Beyond greed and fear: understanding behavioral finance and the
psychology of investing (Boston, Mass.: Harvard Business School Press).

\bibitem{Shillerexu} Shiller, R.J. (2000)
{\it Irrational exuberance} (Princeton University Press, Princeton, NJ: ).

\bibitem{Shleifer} Shleifer, A. (2000)
Inefficient markets: an introduction to behavioral finance
(New York: Oxford University Press).

\bibitem{Siegel} J.J. Siegel,
What Is an Asset Price Bubble? An Operational Definition,
European Financial Management 9 (2003) 11-24.

\bibitem{Sorcrashbook} D. Sornette
Why Stock Markets Crash, Critical Events in Complex Financial Systems
(Princeton University Press, Princeton NJ, 2003).

\bibitem{PR03} D. Sornette,
Critical market crashes, Physics Reports 378(1) (2003) 1-98.

\bibitem{SA02} D. Sornette and J.V. Andersen,
A Nonlinear Super-Exponential Rational Model of Speculative Financial Bubbles,
Int. J. Mod. Phys. C 13 (2002) 171-188.

\bibitem{SJ01QF} D. Sornette and A. Johansen,
Significance of log-periodic precursors to financial crashes,
Quantitative Finance 1(4) (2001) 452-471.

\bibitem{SJB95} D. Sornette, A. Johansen and J.-P. Bouchaud,
Stock market crashes, Precursors and Replicas, J.Phys.I France 6 (1996) 167-175.

\bibitem{STZ03PA} D. Sornette, H. Takayasu and W.-X. Zhou,
Finite-Time Singularity Signature of Hyperinflation, Physica A 325
(2003) 492-506.

\bibitem{SZ02QF} D. Sornette and W.-X. Zhou, The US 2000-2002
Market Descent: How Much Longer and Deeper? Quantitative Finance 2
(2002) 468-481.

\bibitem{WXS03} D. Sornette and W.-X. Zhou,
Predictability of Large Future Changes in Complex Systems,
submitted to the International Journal of Forecasting (2004)
(http://arXiv.org/abs/cond-mat/0304601)

\bibitem{van} J. van Biesebroeck, Robustness of Productivity
Estimates, NBER Working Paper No. W10303 (2004)
(http://ssrn.com/abstract=502889).


\bibitem{Visano} B.S. Visano,
A Socioeconomic Perspective,
The American Journal of Economics and Sociology 61 (2002) 801-827.

\bibitem{Weidlich00} W. Weidlich,
Sociodynamics: A systematic approach to mathematical modelling in the
social sciences (Harwood Academic Publishers, Amsterdam, 2000).

\bibitem{WeidlichHaag} W. Weidlich and G. Haag, Concepts and models of a
quantitative sociology: The dynamics of interacting populations
(Springer, Berlin, 1983)

\bibitem{ZS03PAGlobal} W.-X. Zhou and D. Sornette,
Evidence of a worldwide stock market log-periodic antibubble since
mid-2000, Physica A 330 (2003) 543-583.


\end{thebibliography}
\end{document}